\documentclass[letterpaper]{article}
\usepackage[preprint]{aaai2027}
\usepackage[hyphens]{url}
\usepackage{graphicx}
\usepackage{natbib}
\usepackage{booktabs}
\usepackage{amsmath}
\usepackage{amssymb}
\usepackage{placeins}
\title{Beyond Frequency: Dissonance Spectrum for Perceptually Motivated Music Understanding}
\author{Tianle Wang$^{1,2}$, Xinyi Tong$^{1,2}$, Liangke Zhao$^{1,2}$,\\
Jishang CHEN$^{1,2}$, Sirui Zhang$^{1,2}$, Haoxin Zhang$^{1,2}$,\\
Xin Jin$^{1}$, Duo XU$^{1}$, Xiaobing Li$^{2}$, Song-Chun Zhu$^{1,3}$}
\affiliations{$^{1}$Beijing Institute for General Artificial Intelligence, Beijing, China\\
$^{2}$Central Conservatory of Music, Beijing, China\\
$^{3}$Peking University, Beijing, China}

\begin{document}
\maketitle
-
\begin{abstract}
Conventional music representations describe acoustic energy over time and frequency but do not explicitly expose relations among simultaneous frequency components. We introduce the \emph{Dissonance Spectrum} (DS), a nonnegative time--frequency representation that applies a tolerance-based rational pitch-relation kernel with logarithmic harmonic distance to a constant-Q spectrum and attributes aggregate pairwise interactions back to individual frequency bins. Controlled music-theory tests show strong ordinal agreement for intervals, harmonic-function connections, and church modes, and moderate positive agreement across diverse chord voicings. DS is then encoded by a lightweight parallel branch whose zero-initialized residual projection preserves the baseline function at initialization. Across six paired training seeds in open-ended music question answering and categorical and dimensional music emotion recognition, DS obtains the highest mean on every reported endpoint relative to the unchanged baseline, a parameter-matched Gaussian-input branch, and an architecture-matched magnitude-CQT branch. These results support DS as an interpretable, complementary representation, while listener-specific perception and broader task coverage remain open problems.
\end{abstract}

\refstepcounter{figure}\label{fig:idea}
\begin{center}
\includegraphics[trim=20 80 20 80,clip,width=.90\columnwidth]{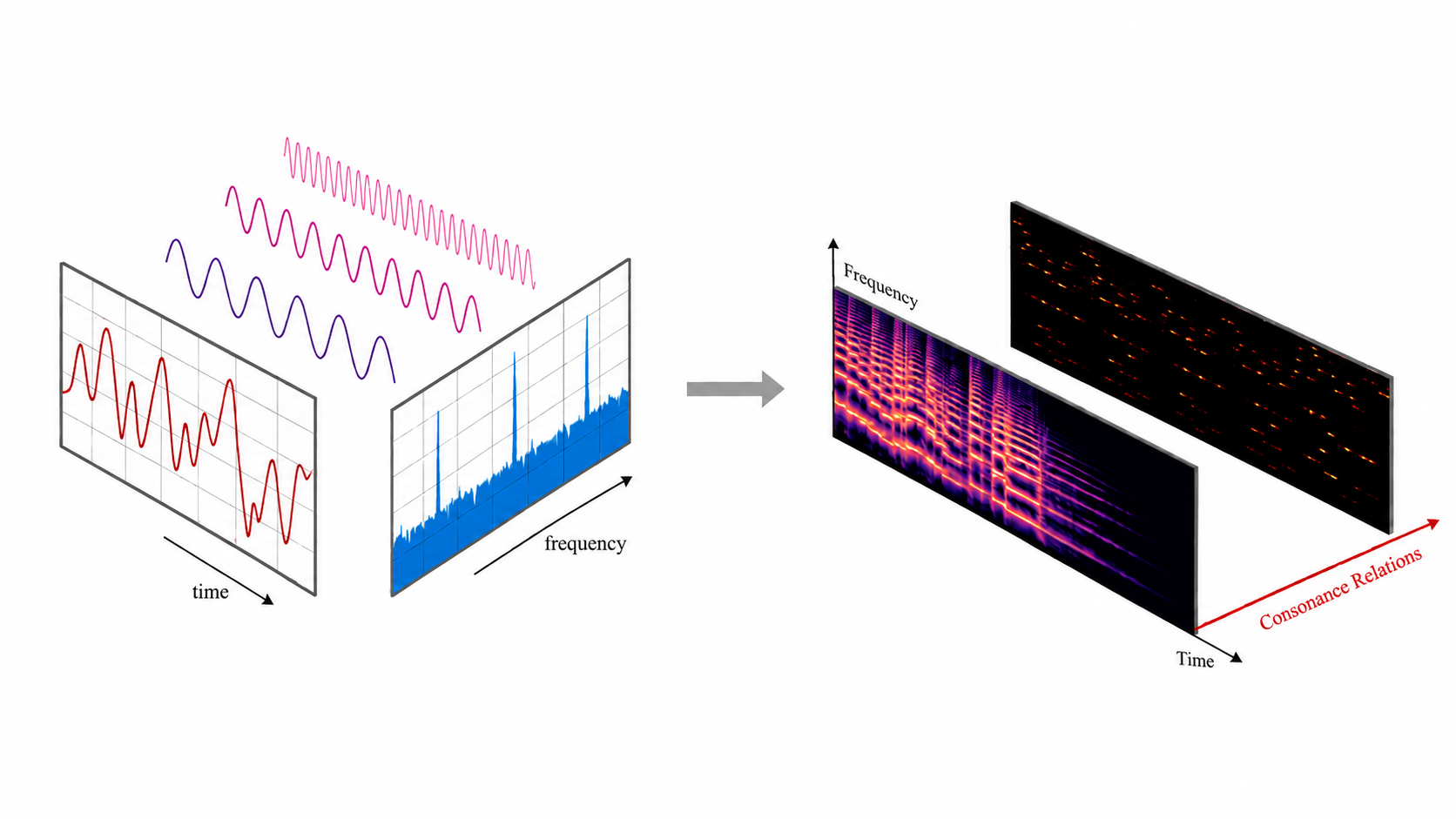}
\end{center}
\vspace{-6pt}
\noindent{\small Figure~\thefigure: DS converts spectral energy into a time--frequency map of modeled consonance--dissonance relations.\par}
\vspace{4pt}

\section{Introduction}
Music understanding extends beyond detecting acoustic events: listeners also respond to relations among simultaneous frequency components. Beating, critical-band interactions, periodicity, and harmonic organization contribute to consonance, dissonance, stability, tension, and emotion \citep{helmholtz1954sensations,plomp1965tonal,langner1992periodicity,stolzenburg2015harmony,harrison2020simultaneous}. Neural systems learn rich musical information from waveforms, time--frequency representations, and audio--text supervision \citep{lu2021spectnt,li2024mert,won2024musicfm,huang2022mulan,elizalde2023clap,liu2024mullama,deng2024musilingo}. Yet spectral magnitude mainly marks active components, while learned embeddings entangle multiple attributes; neither explicitly localizes simultaneous relations under a specified consonance--dissonance model.

Music-specific inductive biases can complement end-to-end learning. Chroma and tonal features expose pitch-class organization, perceptual mid-level attributes connect audio to emotion, MERT uses a CQT teacher, and consonance-aware supervision improves chord estimation \citep{harte2006harmonic,korzeniowski2016feature,aljanaki2018midlevel,chowdhury2019explainable,li2024mert,poltronieri2025discord}. Lightweight mode injection likewise benefits symbolic emotion recognition \citep{xia2026modeguided}. These results motivate a relational representation that makes information already present in the signal easier to inspect and learn.

We introduce the \textbf{Dissonance Spectrum} (Figure~\ref{fig:idea}). DS combines tolerance-based rational approximation \citep{stolzenburg2015harmony} with logarithmic harmonic distance \citep{tenney1984johncage}, applies the resulting kernel across a magnitude CQT, and attributes weighted pair relations back to their time--frequency locations. It is a deterministic reorganization of the signal, not an additional observed modality, and is computed efficiently by frequency-axis correlation.

We validate the operator on intervals, chord qualities, functional chord connections, and scales, then add a lightweight DS encoder to MU-LLaMA for open-ended music question answering and to Music2Emo for categorical and dimensional emotion recognition. Parameter-matched Gaussian and architecture-matched magnitude-CQT branches control for added capacity and a second pitch-resolved pathway. Six paired seeds and exploratory mechanism variants test whether gains are consistent with ordered relational structure rather than input resolution or model size alone.

Our contributions are threefold:
\begin{itemize}
\item We formulate a nonnegative time--frequency DS that localizes amplitude-weighted rational pitch relations and derive an efficient correlation implementation.
\item We establish controlled evidence that the kernel and audio representation reproduce predefined ordinal trends across intervals, chord qualities, functional connections, and modes.
\item We design lightweight plug-and-play adapters and show that DS achieves higher six-seed mean performance than the baseline, a parameter-matched Gaussian branch, and an architecture-matched CQT branch in two model families.
\end{itemize}

\section{Related Work}

\paragraph{Music representations and perceptual priors.}
The CQT provides a logarithmic frequency axis aligned with musical pitch, chroma folds energy into pitch classes, and tonal-space descriptors encode harmonic proximity \citep{brown1991cqt,harte2006harmonic,mueller2011chroma}. Neural models learn broader musical attributes: SpecTNT separates spectral and temporal attention, MERT and MusicFM provide transferable music embeddings, and MuLan, CLAP, MU-LLaMA, and MusiLingo connect audio representations to language \citep{lu2021spectnt,li2024mert,won2024musicfm,huang2022mulan,elizalde2023clap,liu2024mullama,deng2024musilingo}. These representations are effective for downstream tasks but do not provide a stable, directly inspectable account of local consonance or dissonance. Recent human-written music-QA evaluation further emphasizes robustness to unimodal shortcuts, motivating cautious interpretation of automatically generated references and lexical-overlap metrics \citep{weck2026hummusqa}.

Music-specific priors remain complementary to learned representations. Chroma-based networks improve chord recognition; listener-rated mid-level attributes bridge audio and emotion; MERT uses a CQT teacher; and consonance-aware distances and label smoothing improve chord estimation \citep{korzeniowski2016feature,aljanaki2018midlevel,chowdhury2019explainable,li2024mert,poltronieri2025discord}. DS follows this knowledge-guided direction but is computed continuously from audio and preserves time--frequency attribution.

\paragraph{Consonance perception and computational models.}
Consonance and dissonance denote related but nonidentical acoustic, perceptual, and music-theoretical concepts \citep{cazden1980definition,wand2012conception}. Interference accounts relate sensory dissonance to beating among nearby partials and auditory critical bands \citep{helmholtz1954sensations,plomp1965tonal,hutchinson1978acoustic}. Periodicity and harmonicity accounts associate consonance with compact common periods, virtual fundamentals, or harmonic templates \citep{langner1992periodicity,parncutt1989harmony,stolzenburg2015harmony}. Many implementations reduce these relations to a scalar, which supports ranking but does not identify the frequency regions responsible for the value.

Timbre can reshape consonance curves by changing partial locations and amplitudes \citep{sethares2005tuning,marjieh2024timbral}. Comparative modeling and listener studies further indicate that roughness, periodicity, spectral structure, experience, register, and listener characteristics all contribute to consonance judgments \citep{cousineau2012basis,harrison2020simultaneous,eerola2021anatomy,mcdermott2016indifference,kaklamani2026dyads}. We therefore treat DS as a perceptually motivated low-level cue rather than a complete model of musical preference.

\paragraph{Recorded music and consonance-aware music AI.}
Applying dissonance models to recordings is difficult because real audio contains overlapping sources, transients, noise, tuning variation, and time-varying balance. \citet{schwaer2025measuring} estimate time-varying sensory dissonance in multitrack recordings and attribute mixture-level dissonance to tracks. Scalar roughness descriptors and learned mid-level dissonance ratings have also supported music emotion recognition \citep{aljanaki2018midlevel,panda2023audiofeatures}. DS instead attributes aggregate relations to time--frequency bins, requires no stems or chord labels, and can be encoded as an additional input to pretrained music systems.

\section{Method}
\label{sec:method}

\subsection{Overview}
Our method has two parts. First, the \emph{Dissonance Spectrum} (DS) assigns each active CQT bin its amplitude-weighted relation to the other active components. A continuous kernel combines tolerance-based rational approximation \citep{stolzenburg2015harmony} with Tenney's harmonic distance \citep{tenney1984johncage}; sampling it on the CQT grid permits efficient one-dimensional frequency correlation instead of a $K\times K\times T$ pair tensor. Second, a lightweight encoder and shape-preserving gated residual adapter inject cached DS maps into a host representation without changing the host outputs, heads, or losses. Zero initialization preserves the pretrained baseline function at the start of fine-tuning. We next define intrinsic and cross-reference DS, derive the correlation form, and describe both host integrations; full index derivations and optional temporal variants are in the supplement.

\subsection{Pitch-Relation Kernel}
\paragraph{Rational candidates and complexity.}
For a maximum numerator and denominator $Q$, the reduced rational candidate set is
\begin{equation}
\mathcal{R}_Q
=
\left\{
\frac{p}{q}
\;\middle|\;
1 \le p \le q \le Q,\;
p,q \in \mathbb{Z}^{+},\;
\gcd(p,q)=1
\right\}.
\label{eq:rational-set}
\end{equation}
For a reduced ratio, we use the complexity function
\begin{equation}
C\left(\frac{p}{q}\right)=\log_{2}(pq),
\label{eq:complexity}
\end{equation}
which is the logarithmic product form of Tenney's harmonic distance \citep{tenney1984johncage}. Here $Q$ is a finite search-resolution hyperparameter rather than a direct estimate of a listener attribute. For visualization only, simplicity is the reversed complexity scale,
\begin{equation}
S\left(\frac{p}{q}\right)
=
\max_{r \in \mathcal{R}_Q} C(r)
-
C\left(\frac{p}{q}\right).
\end{equation}

Given two frequencies, we order their ratio as
\begin{equation}
r = \frac{\min(f_1,f_2)}{\max(f_1,f_2)},
\end{equation}
and define the relative-error neighborhood
\begin{equation}
\mathcal{N}_\alpha(r) = \left\{ x \in \mathbb{R}^+ \;\middle|\; |x - r| \le \alpha r \right\}.
\label{eq:tolerance}
\end{equation}
We use $\alpha=.01$, following the 1\% relative tolerance used in Stolzenburg's smoothed periodicity formulation \citep{stolzenburg2015harmony}. Let
\begin{equation}
\frac{p^{*}}{q^{*}}
=
\arg\min_{\frac{p}{q} \in \mathcal{R}_Q}
\left|
r-\frac{p}{q}
\right|.
\end{equation}
The pairwise value is then
{\small
\begin{equation}
D(f_1,f_2)
=
\begin{cases}
\displaystyle
\min_{\frac{p}{q} \in \mathcal{R}_Q \cap \mathcal{N}_{\alpha}(r)}
C\left(\frac{p}{q}\right),
&
\text{if }
\mathcal{R}_Q \cap \mathcal{N}_{\alpha}(r)
\neq \emptyset,
\\[10pt]
\displaystyle
C\left(\frac{p^{*}}{q^{*}}\right),
&
\text{otherwise}.
\end{cases}
\label{eq:pair-kernel}
\end{equation}
}
Thus ratios admitting a simpler approximation inside the tolerance receive a lower value; otherwise the closest candidate is used. This is a periodicity/harmonic-distance cue, not a critical-band roughness model.

\paragraph{Continuous pitch intervals and octave folding.}
We map pitch to frequency by
\begin{equation}
\mathrm{freq}(\mathrm{pitch}) = 440 \cdot 2^{\frac{\mathrm{pitch} - 69}{12}}
\end{equation}
and evaluate
\begin{equation}
D_{\mathrm{interval}}(\Delta p)
=
D\big(
\mathrm{freq}(\mathrm{pitch}_{\mathrm{ref}}+\Delta p),\,
\mathrm{freq}(\mathrm{pitch}_{\mathrm{ref}})
\big).
\end{equation}
With
\begin{equation}
D_{\mathrm{int,max}}
=
\max_{\Delta p \in [0,12)}
D_{\mathrm{interval}}(\Delta p),
\end{equation}
for $\Delta p\in[0,12)$, the normalized function is
\begin{equation}
D_{\mathrm{norm}}(\Delta p)
=
\frac{D_{\mathrm{interval}}(\Delta p)}{D_{\mathrm{int,max}}}
\in [0,1].
\end{equation}
Let $\delta_{12}(\Delta p)=|\Delta p|\bmod 12\in[0,12)$ denote the
octave-folded interval magnitude. The resulting even function is
\begin{equation}
\bar{D}(\Delta p)
=
D_{\mathrm{norm}}\!\left(\delta_{12}(\Delta p)\right),
\qquad \bar D(0)=0.
\label{eq:folded-kernel}
\end{equation}
The continuous pitch argument allows evaluation between equal-tempered bins, while octave folding implements a pitch-chroma assumption. Because pitch height and pitch chroma can both affect perception \citep{wagner2022pitchchroma}, folding is an explicit modeling choice rather than a claim that register never matters. The resulting curve is shown in the supplement.

\subsection{Intrinsic Dissonance Spectrum}
\paragraph{CQT representation and preprocessing.}
Let $\mathbf X\in\mathbb R_{\ge0}^{K\times T}$ be a magnitude CQT and $\vec x^{\,t}=\mathbf X(:,t)$. With $B$ bins per octave, its pitch and frequency grids are
\begin{equation}
\vec{p}
=
\left[
\mathrm{pitch}_{\min}+\frac{12k}{B}
\right]_{k=0}^{K-1},
\qquad
\vec{f}
=
\left[
\mathrm{freq}(\mathrm{pitch}_k)
\right]_{k=0}^{K-1}.
\end{equation}
Values below a fixed floor are set to zero, and a local spectral-peak mask may suppress nonpeak bins. We use excerpt-level global-maximum normalization in the downstream experiments and the context-dependent normalization specified for the controlled validation. DS and its magnitude-CQT control always share the same normalization within a comparison.

\paragraph{Amplitude-weighted attribution.}
Motivated by the pairwise aggregation of spectral partials in dissonance-curve models \citep{sethares1994adaptive,sethares2005tuning}, we use the simple amplitude-product coefficient and pair value
\begin{equation}
A(x_1,x_2)=x_1x_2,
\qquad
\bar{D}(\Delta p;\,x_1,x_2)=x_1x_2\bar{D}(\Delta p).
\end{equation}
We define the intrinsic DS element at bin $k$ and frame $t$ as
\begin{equation}
 d^t_k
=
\frac{1}{K}
\sum_{l=0}^{K-1}
x^t_k\,x^t_l\,\bar{D}(p_l-p_k),
\label{eq:intrinsic-ds}
\end{equation}
and collect the frame vectors as
\begin{equation}
\mathbf D=
\begin{bmatrix}
\vec d^{\,0}&\cdots&\vec d^{\,T-1}
\end{bmatrix}
\in\mathbb R^{K\times T}.
\end{equation}
The factor $1/K$ averages the $K$ reference-bin contributions. Equation~\ref{eq:intrinsic-ds} differs from a frame-level scalar: every pair contributes to both participating locations through their own target factors, so the map preserves where the modeled relations occur. All terms are nonnegative; inactive bins have zero attribution.

\subsection{Cross-Reference Dissonance Spectrum}
\begin{figure*}[t]
\centering
\includegraphics[width=.42\textwidth]{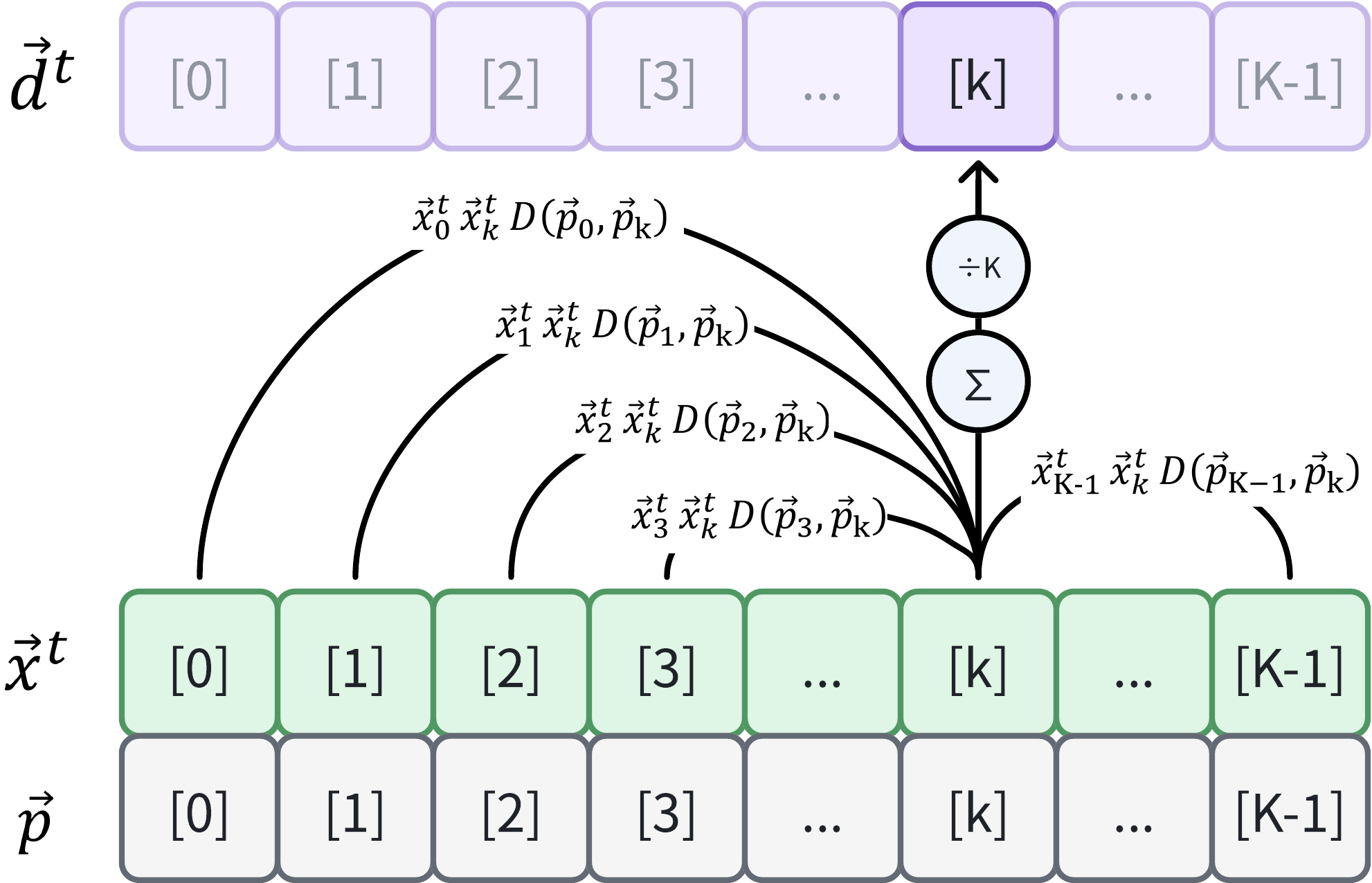}\hfill
\includegraphics[width=.32\textwidth]{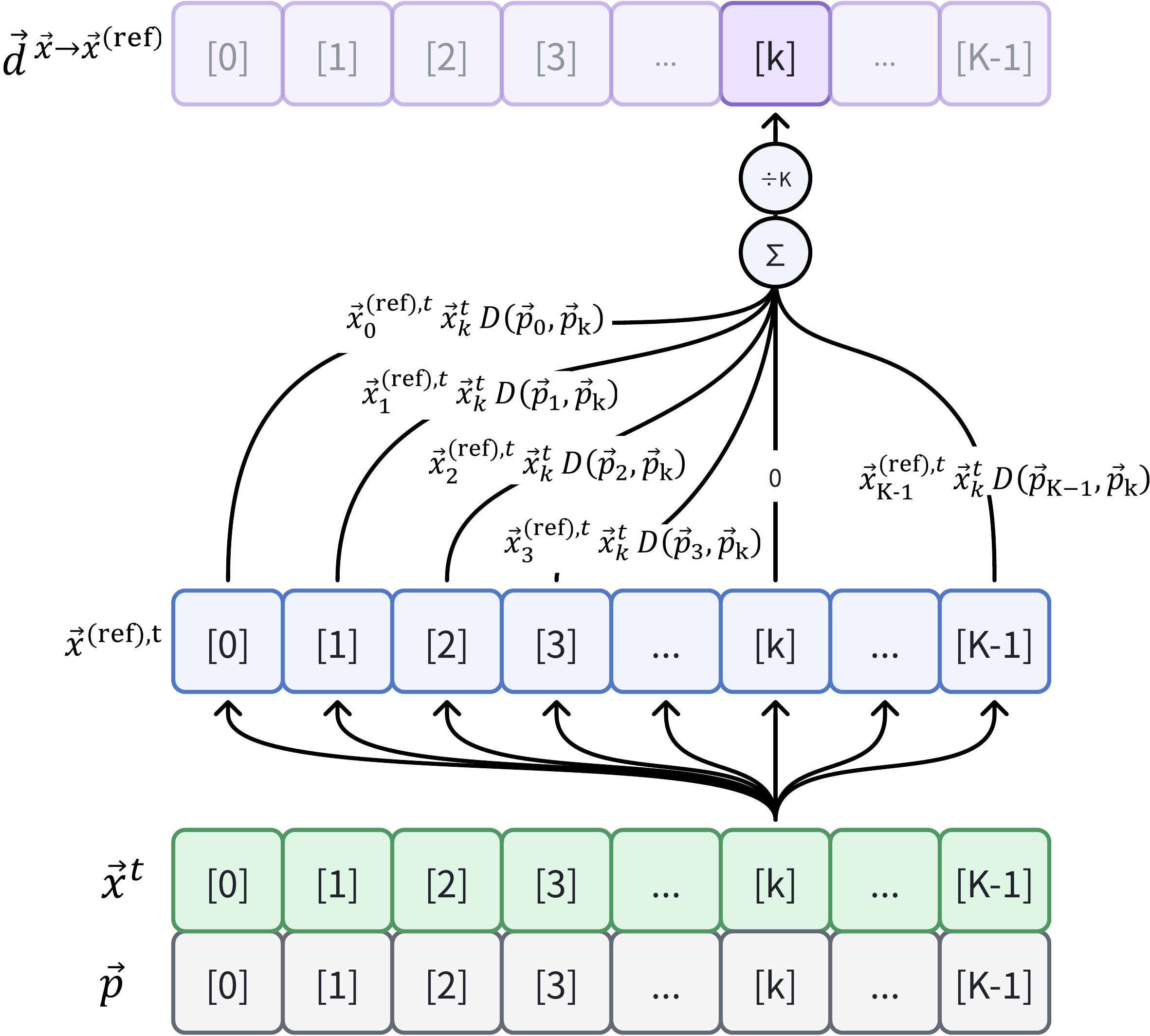}
\caption{Conceptual DS attribution. Left: intrinsic DS aggregates pairwise relations within one spectrum. Right: cross-reference DS evaluates the target spectrum against a separate reference while retaining target-bin attribution.}
\label{fig:intrinsic-cross-concepts}
\end{figure*}

The same operator can attribute the relation between a target spectrum $\vec{x}$ and a separate reference spectrum $\vec{x}^{(\mathrm{ref})}$. In vector form,
\begin{equation}
\vec{d}
^{\,\vec{x} \to \vec{x}^{(\mathrm{ref})}}
=
\frac{1}{K}
\left(
\vec{\mathcal{D}^\pm}
\star_{\mathrm{valid},f}
\vec{x}^{(\mathrm{ref})}
\right)
\odot\vec{x}.
\label{eq:cross-ds}
\end{equation}
Here $\star_{\mathrm{valid},f}$ denotes the target-aligned valid frequency-axis correlation defined below. Intrinsic DS is the special case $\vec{x}^{(\mathrm{ref})}=\vec{x}$. A fixed tonic, chord, or spectral template can instead provide an explicit harmonic reference while the final factor $\vec{x}$ preserves attribution to the target bins. Figure~\ref{fig:intrinsic-cross-concepts} contrasts the two constructions.

\subsection{Efficient Correlation Form}
\begin{figure}[t]
\centering
\includegraphics[width=.98\columnwidth]{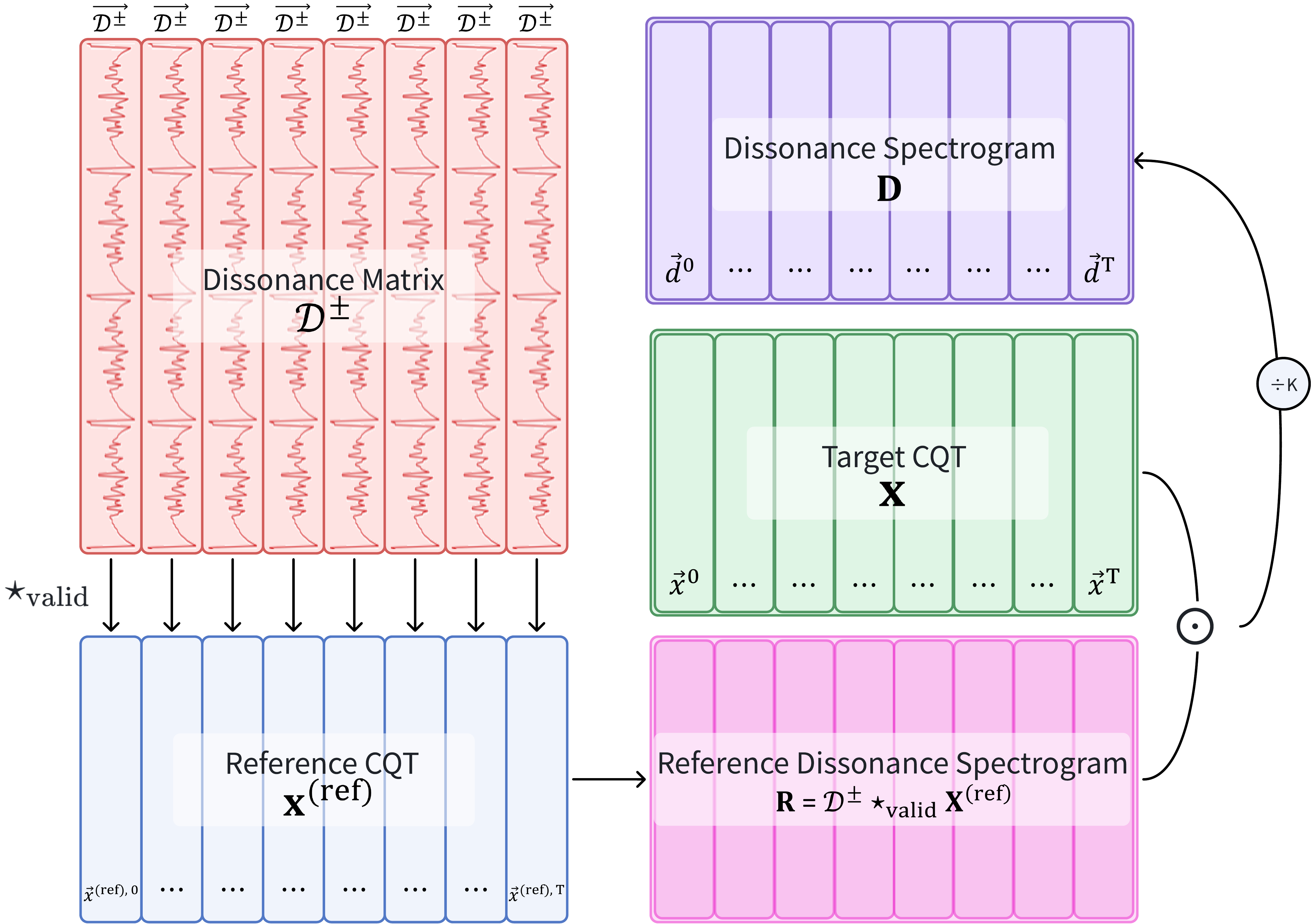}
\caption{Frequency-axis matrix implementation of cross-reference DS. Intrinsic DS follows by setting the reference and target CQT matrices equal.}
\label{fig:matrix-compute}
\end{figure}

Because $\bar D$ depends only on pitch difference, the $K^2$ pair table need not be stored. Define the length-$(2K-1)$ sampled kernel
\begin{equation}
\begin{aligned}
\mathcal D^\pm_m&=\bar D\!\left(\frac{12m}{B}\right),\\[-2pt]
m&=-(K-1),\ldots,K-1,
\qquad
\vec{\mathcal D^\pm}\in\mathbb R^{2K-1}.
\end{aligned}
\end{equation}
stored in increasing $m$ order, with center $\bar D(0)=0$. The target-aligned index convention is given in the supplement. Since
\begin{equation}
\bar{D}(p_l-p_k)
=
\vec{\mathcal{D}^{\pm}}_{K+l-k-1},
\end{equation}
Equation~\ref{eq:intrinsic-ds} becomes
\begin{equation}
\vec{d}^{\,t}
=
\frac{1}{K}
\left(
\vec{\mathcal{D}^\pm}
\star_{\mathrm{valid},f}
\vec{x}^{\,t}
\right)
\odot
\vec{x}^{\,t}.
\label{eq:corr-ds}
\end{equation}
For all frames, repeat the same kernel across time as
\begin{equation}
\mathbf{\mathcal{D}}^{\pm}
=
\begin{bmatrix}
\vec{\mathcal{D}}^{\pm}&\vec{\mathcal{D}}^{\pm}&\cdots&\vec{\mathcal{D}}^{\pm}
\end{bmatrix}
\in\mathbb{R}^{(2K-1)\times T},
\end{equation}
and compute
\begin{equation}
\mathbf{D}
=
\frac{1}{K}\,
(\mathbf{\mathcal{D}}^{\pm}
\star_{\mathrm{valid},f}
\mathbf{X})\odot\mathbf{X}.
\label{eq:matrix-ds}
\end{equation}
The correlation is one-dimensional along frequency and returns $K\times T$ values. It avoids the $O(K^2T)$ intermediate storage of direct pairwise attribution and can be batched with standard correlation operators. Figure~\ref{fig:matrix-compute} shows the corresponding matrix computation; the full index derivation and optional temporal variants are included in the supplement.

\subsection{Representation Properties and Extraction}
\paragraph{Local attribution rather than a scalar score.}
A conventional spectrum reports energy at $(k,t)$; DS reports how strongly that component participates in the modeled relations of its frame. Unlike a scalar score that collapses register and can only modulate a representation globally, the $K\times T$ map preserves register and instrumentation cues. A downstream encoder can therefore distinguish a narrow high-valued interaction among upper partials from a broad interaction spanning the bass and midrange. Each high value coincides with nonzero target magnitude and can be traced to the reference components selected by the shifted kernel.

\paragraph{Nonnegativity, symmetry, and translation structure.}
Nonnegative magnitudes and kernel values make DS nonnegative. The ordered ratio and octave-folded difference make the pair relation symmetric, while the final target factor restores bin-specific attribution. Because coefficients depend on pitch difference rather than absolute bin index, equal intervals share coefficients across register. This translation structure permits cross-correlation and differs from an unconstrained learned two-dimensional filter: before training, the same specified relation is treated consistently throughout the CQT range, subject to octave folding.

\paragraph{Continuous pitch grid.}
The observed frequency ratio is not quantized to the rational candidate set. Equation~\ref{eq:pair-kernel} selects a candidate within tolerance or the nearest candidate, and Equation~\ref{eq:folded-kernel} is sampled at $12/B$-semitone CQT increments. DS therefore responds to detuning, expressive variation, and inharmonic partials rather than only twelve integer pitch classes. The supplement gives qualitative microtonal and timbral examples without assigning a perceptual ranking.

\paragraph{Offline extraction procedure.}
For each excerpt, we compute magnitude CQT with the shared floor, peak mask, and normalization; sample $\vec{\mathcal D}^{\pm}$ from $Q$, $\alpha$, and the CQT grid; and evaluate Equation~\ref{eq:matrix-ds}. Finite outputs are transformed by $\log(1+\mathbf D)$ and cached with the audio hash, sample rate, hop, pitch range, $B$, $Q$, $\alpha$, and preprocessing flags. Training retrieves the DS segment at the baseline branch's temporal indices, with deterministic validation and test crops. This avoids repeated spectral computation and ensures that the baseline and DS branches observe the same recording portion.

\subsection{Plug-and-Play Augmentation of Music Understanding Models}
\label{sec:plugin}
DS supplements rather than replaces the host waveform encoder or task representation. Let $\mathbf H^{(l)}\in\mathbb R^{N\times d}$ be the hidden sequence, or a pooled token when $N=1$, at insertion layer $l$. A lightweight encoder maps the cached feature to temporal tokens,
\begin{equation}
\mathbf Z=E_{\phi}\!\left(\log(1+\mathbf D)\right)
\in\mathbb R^{M\times d_s},
\end{equation}
using convolutional frequency compression followed by temporal convolution or attention. DS tokens share the baseline branch's excerpt indices and valid-frame mask.

A shape-preserving adapter lets the module attach to host representations of different dimensions. The baseline tokens query the DS tokens and receive a gated residual update:
\begin{align}
\mathbf C
&=\operatorname{softmax}\!\left(
\frac{(\mathbf H^{(l)}\mathbf W_Q)(\mathbf Z\mathbf W_K)^{\top}}
{\sqrt{d_a}}
\right)(\mathbf Z\mathbf W_V),
\label{eq:ds-cross-attn}\\
\mathbf G
&=\sigma\!\left(\mathbf W_G[\mathbf H^{(l)};\mathbf C]+\mathbf b_G\right),\\
\widetilde{\mathbf H}^{(l)}
&=\mathbf H^{(l)}+\mathbf G\odot(\mathbf C\mathbf W_O).
\label{eq:ds-gated-residual}
\end{align}
When $N=1$, the same equations give a pooled adapter. The unchanged output shape preserves the decoder, heads, losses, and evaluation. We initialize $\mathbf W_O$ to zero and $\mathbf b_G$ negatively, so $\widetilde{\mathbf H}^{(l)}=\mathbf H^{(l)}$ initially. Disabling the branch recovers the baseline checkpoint without conversion. Independent extraction, shape-preserving insertion, and exact fallback define the plug-and-play property.

\begin{figure*}[t]
\centering
\includegraphics[width=.88\textwidth]{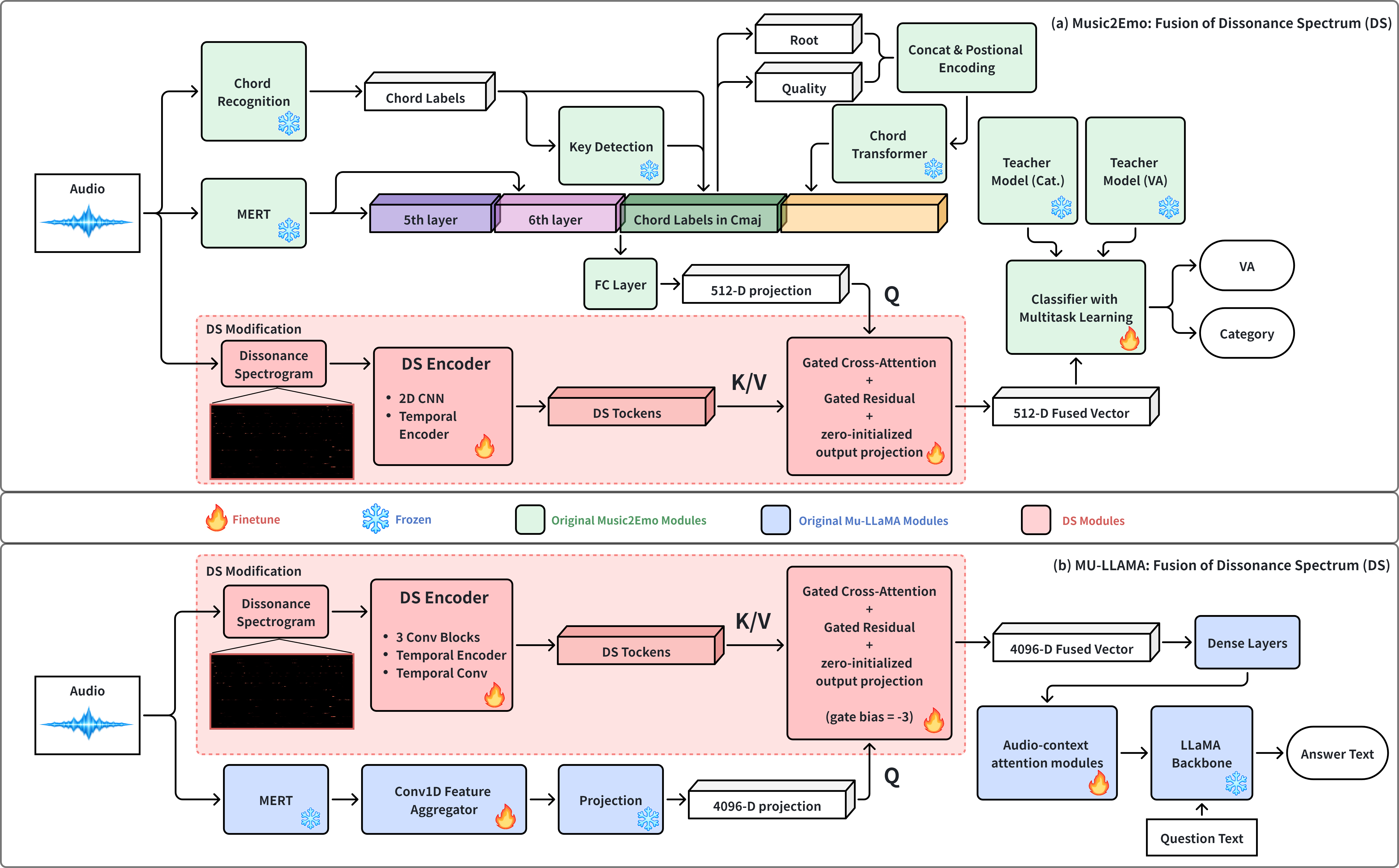}
\caption{Plug-and-play DS integration in Music2Emo (top) and MU-LLaMA (bottom). The DS branch provides key/value tokens to a gated residual adapter while the original projection provides the query; snowflakes and flames mark frozen and fine-tuned modules.}
\label{fig:model-overall}
\end{figure*}

For MU-LLaMA, the 576-bin map is pooled to 144 channels and processed by three convolutional blocks, a 128-dimensional temporal encoder, depthwise temporal convolution, and single-head attention. The pooled adapter follows the original audio projection while MERT and LLaMA remain frozen. For Music2Emo, temporal DS tokens are queried after the unchanged 512-dimensional projection of MERT, chord, and key-mode features, before the original emotion heads. Figure~\ref{fig:model-overall} shows both insertion paths and frozen versus trainable modules.

To control for adding capacity or a second pitch-resolved pathway, the CQT and Gaussian conditions reuse the same encoder, fusion point, output dimension, and trainable-parameter budget. They replace $\mathbf D$ with the aligned normalized magnitude CQT or a fixed random input, respectively.

\paragraph{Interpretive scope.}
DS operationalizes one periodicity/harmonic-distance-inspired relation and its amplitude-weighted localization. It does not explicitly model auditory-filter bandwidths, masking, learned tonal syntax, or cultural preference. The task experiments therefore test whether this representation is useful as an inductive bias, not whether it is a complete perceptual theory of consonance.

\providecommand{\rmacro}{\overline{R^2}_{\mathrm{VA}}}
\section{Experiments}
\label{sec:experiments}

\subsection{Experimental Setup}
\paragraph{Implementation, compute, and reproducibility.}
Controlled calculations are deterministic and parameter-free. PyTorch downstream systems run on a Slurm-managed Linux cluster with one NVIDIA A100 per job and no multi-GPU parallelism. CQT and DS features are extracted once per exact excerpt, cached with metadata, and reused across conditions; the submitted environment lock records software and pretrained-model revisions.

We use paired seeds $\{17,42,101,2025,2026,3407\}$. Within each dataset and seed, all four conditions share splits, excerpts, masks, minibatch order, optimization, early stopping, checkpoint selection, and evaluation; random generators use the listed seed. Except for PMEmo's released chorus clips, inputs are 24-kHz, 45-second excerpts with zero padding or one deterministic crop shared by all branches. CQT uses a 1,024-sample hop, eight octaves, 72 bins per octave, $f_{\min}=\mathrm{C1}$, and 576 bins; DS applies the proposed transform and $\log(1+\mathbf D)$. Primary endpoints are BERTScore-R for MusicQA and $\rmacro$, the mean of six valence/arousal $R^2$ values, for Music2Emo. Tests use unrounded seed-level values.

\subsection{Controlled Music-Theory Validation}
\label{sec:theory-validation}
We first test whether the relation operator recovers controlled musical structure from rendered piano audio: 13 dyads from unison through the octave, four representative and 13 extended chord voicings, seven diatonic connections to C major, and seven church modes. CQT uses 22.05 kHz audio, four frames per second, eight octaves, 72 bins per octave, $f_{\min}=\mathrm{C1}$, and $K=576$; the rational search uses $Q=60$ and $\alpha=.01$. Intervals and scales use a C4 reference, chord qualities equally average tonic-reference and intrinsic DS, and connections use C major. These settings test the same operator under phenomenon-specific reference contexts; downstream models use intrinsic DS.

Interval, chord, and connection maps are reduced to $D_{\max}=\max_t\sum_k \mathbf D(k,t)$; sequential scales use $D_{\mathrm{sum}}=\sum_t\sum_k\mathbf D(k,t)$. Spearman $\rho$ and Kendall $\tau_b$ compare these summaries with predefined orders. The interval order broadly agrees with classic dyad studies \citep{malmberg1918consonance,schwartz2003statistical}; chord, function, and mode orders are theory-derived hypotheses, so they test internal music-theoretical consistency rather than population-level perception.

\begin{table}[t]
\centering
\scriptsize
\setlength{\tabcolsep}{2.5pt}
\caption{Controlled validation. Parentheses give two-sided $p$-values; the four-class row is an ordering check.}
\label{tab:theory-validation}
\resizebox{\columnwidth}{!}{%
\begin{tabular}{lccc}
\toprule
Test & $n$ & Spearman $\rho$ & Kendall $\tau_b$ \\
\midrule
Intervals & 13 & .951 ($6.36\!\times\!10^{-7}$) & .846 ($5.20\!\times\!10^{-6}$) \\
Chord quality (4 exemplars) & 4 & 1.000 (--) & 1.000 (--) \\
Chord quality (13 voicings) & 13 & .626 (.022) & .462 (.030) \\
Functional connections & 7 & .794 (.033) & .655 (.054) \\
Church modes & 7 & .893 (.0068) & .810 (.0107) \\
\bottomrule
\end{tabular}}
\end{table}

\begin{figure}[t]
\centering
\includegraphics[width=.96\columnwidth]{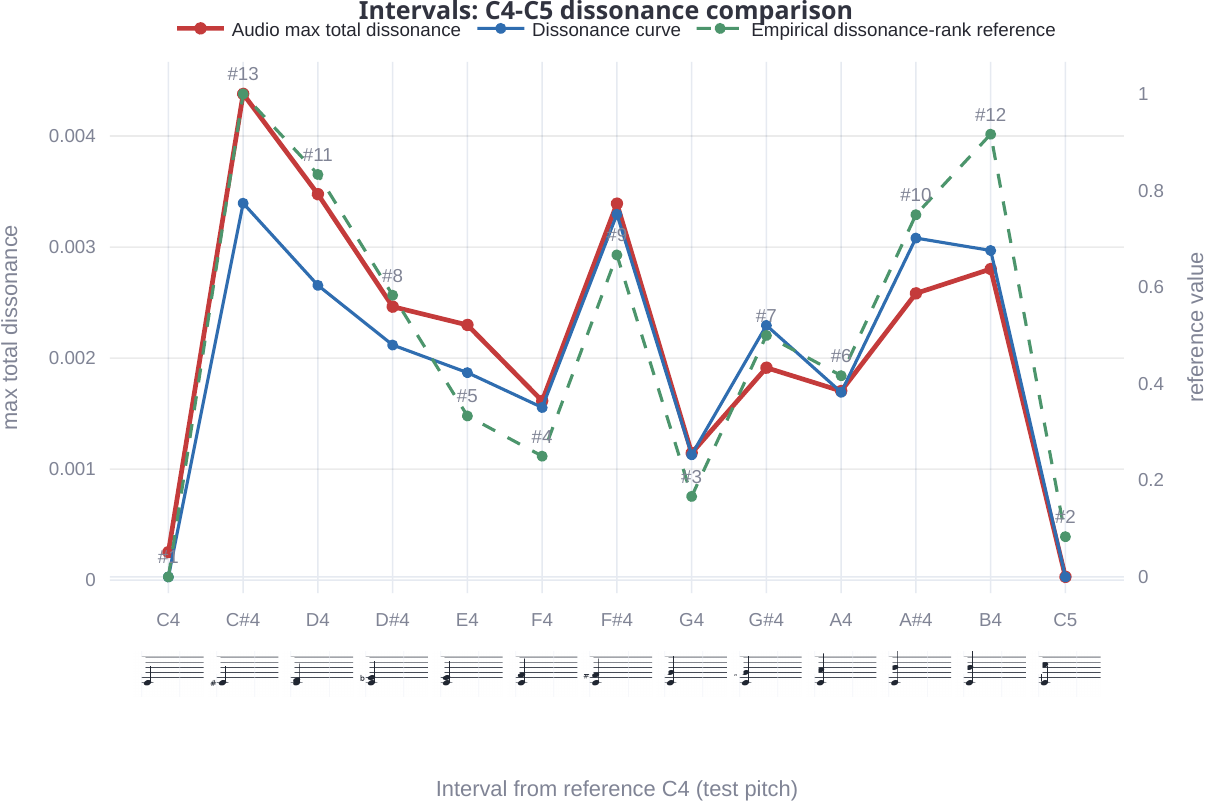}
\caption{Interval validation: audio DS maxima align with the sampled dissonance curve and the predefined rank reference.}
\label{fig:interval-main}
\end{figure}

Intervals show the strongest agreement: C--C$\sharp$ is maximal, the tritone is high, and the perfect fifth and octave are among the lowest (Figure~\ref{fig:interval-main}). Audio DS maxima track both the sampled curve and predefined ranks. Four chord exemplars follow major $<$ minor $<$ suspended $<$ diminished; across 13 voicings the positive but nonmonotonic association reflects spacing and inversion rather than chord labels alone. Tonic-function connections are generally below predominant and dominant connections, while Ionian is near the low end and Locrian highest among modes, with local reversals. Item values, spectral maps, loudness sensitivity, timbre, and microtonal analyses are in the supplement.

\subsection{Music Question Answering}
\paragraph{Protocol.}
MU-LLaMA is fine-tuned on 70,011 question--answer pairs from 7,779 tracks and evaluated on 5,040 pairs from 560 audio-disjoint MTG-Jamendo tracks. All conditions use frozen LLaMA-2 7B and MERT-v1-330M with the released initialization \citep{touvron2023llama2,li2024mert,liu2024mullama}. The Baseline has 4.21M trainable parameters; each matched branch has 5.57M. BF16 AdamW uses gradient accumulation to an effective batch of 32, at most four epochs, and validation-loss early stopping; the best checkpoint is evaluated once on the held-out set. One fixed evaluator reports BLEU, METEOR, ROUGE-L, BERTScore-R, loss, and perplexity \citep{papineni2002bleu,banerjee2005meteor,lin2004rouge,zhang2020bertscore}; exact tokenization, generation, and package settings are in the supplement.

\begin{table}[t]
\centering
\scriptsize
\setlength{\tabcolsep}{2.7pt}
\caption{MusicQA results over six paired training seeds (mean$\pm$SD). $\Delta$ is DS minus Baseline.}
\label{tab:qa}
\resizebox{\columnwidth}{!}{
\begin{tabular}{lccccc}
\toprule
Metric & Baseline & Gaussian & CQT & DS & $\Delta$ \\
\midrule
BLEU $\uparrow$ & .2987$\pm$.0019 & .2985$\pm$.0019 & .3056$\pm$.0015 & \textbf{.3074$\pm$.0014} & $+.0087$ \\
METEOR $\uparrow$ & .3761$\pm$.0017 & .3759$\pm$.0018 & .3838$\pm$.0014 & \textbf{.3857$\pm$.0013} & $+.0096$ \\
ROUGE-L $\uparrow$ & .4556$\pm$.0018 & .4554$\pm$.0019 & .4643$\pm$.0016 & \textbf{.4671$\pm$.0015} & $+.0115$ \\
BERTScore-R $\uparrow$ & .8952$\pm$.0007 & .8950$\pm$.0006 & .8996$\pm$.0006 & \textbf{.9024$\pm$.0015} & $+.0072$ \\
Test loss $\downarrow$ & .625$\pm$.004 & .626$\pm$.004 & .607$\pm$.003 & \textbf{.600$\pm$.002} & $-.025$ \\
Perplexity $\downarrow$ & 1.868$\pm$.008 & 1.870$\pm$.008 & 1.836$\pm$.005 & \textbf{1.822$\pm$.004} & $-.046$ \\
\bottomrule
\end{tabular}}
\end{table}

DS has the highest six-seed mean on every MusicQA metric. BERTScore-R increases by $.0072$ over Baseline, $.0074$ over Gaussian, and $.0028$ over CQT, with positive paired differences for all six seeds. The largest Holm-adjusted paired-$t$ $p$-value is $.0017$, but the exact two-sided sign test gives $.03125$ per comparison and $.09375$ after Holm correction across the three DS-versus-control comparisons. We therefore emphasize repeated-seed direction and effect size, not familywise distribution-free significance. Text metrics measure reference similarity rather than factual musical understanding.

\subsection{Music Emotion Recognition}
\paragraph{Model and data.}
Music2Emo projects MERT layers five and six, chord features, and key mode to 512 dimensions before the DS adapter \citep{kang2025unifiedmusicemotionrecognition}. The Baseline task network has 1.07M trainable parameters; each matched branch adds 0.21M while the 95M-parameter MERT remains frozen. We retain the official MTG-Jamendo split \citep{bogdanov2019mtg} and fixed 70/15/15 track splits for DEAM, EmoMusic, and PMEmo \citep{aljanaki2017deam,soleymani2013songs,zhang2018pmemo,kang2025unifiedmusicemotionrecognition}. Weighted binary cross-entropy covers 56 tags, and mean squared error covers valence and arousal. All conditions share the knowledge-distillation objective, Adam at $10^{-4}$, early stopping, and checkpoint rule; details are in the supplement.

\begin{table}[t]
\centering
\scriptsize
\setlength{\tabcolsep}{3.2pt}
\caption{Music2Emo results over six paired seeds. J-PR/J-ROC are macro tag averages; V/A denote valence/arousal $R^2$. Only $\rmacro$ reports mean$\pm$SD.}
\label{tab:emotion}
\resizebox{\columnwidth}{!}{
\begin{tabular}{lcccc}
\toprule
Metric & Baseline & Gaussian & CQT & DS \\
\midrule
J-PR & $.1539$ & $.1537$ & $.1564$ & \textbf{.1580} \\
J-ROC & $.7806$ & $.7801$ & $.7828$ & \textbf{.7841} \\
DEAM-V & $.5169$ & $.5164$ & $.5272$ & \textbf{.5355} \\
DEAM-A & $.6209$ & $.6202$ & $.6260$ & \textbf{.6291} \\
Emo-V & $.6487$ & $.6479$ & $.6575$ & \textbf{.6642} \\
Emo-A & $.7598$ & $.7590$ & $.7642$ & \textbf{.7668} \\
PM-V & $.5451$ & $.5445$ & $.5532$ & \textbf{.5587} \\
PM-A & $.7926$ & $.7920$ & $.7970$ & \textbf{.7992} \\
$\rmacro$ & .6473$\pm$.0014 & .6467$\pm$.0014 & .6542$\pm$.0016 & \textbf{.6589$\pm$.0018} \\
\bottomrule
\end{tabular}}
\end{table}

DS has the highest mean on every Music2Emo endpoint. $\rmacro$ increases by $.0116$ over Baseline, $.0122$ over Gaussian, and $.0047$ over CQT, with positive differences for all six seeds. The Holm-adjusted exact sign-test value is $.09375$, so we again emphasize direction and effect size. Average valence $R^2$ rises from $.5702$ to $.5861$, and arousal $R^2$ from $.7244$ to $.7317$. CQT is second; the further DS gain is consistent with useful pairwise organization, although compression and normalization differences mean the comparison does not isolate the relation transform alone.

\FloatBarrier

\section{Conclusion}
This work addresses a gap between energy-based spectra and latent learned representations by making one class of simultaneous frequency relations explicit. As a spectrogram reorganizes a waveform without adding a new observation, DS reorganizes magnitude information so that modeled pair relations become directly visible to both researchers and downstream encoders. It applies a continuous, tolerance-based harmonic-distance kernel to magnitude CQT, localizes amplitude-weighted pair relations in time and frequency, and avoids a quadratic pair tensor through frequency-axis correlation. A shape-preserving, zero-initialized adapter then adds this representation to existing systems without changing their output interfaces or baseline function at initialization. Controlled tests recovered strong ordinal agreement for intervals, functional connections, and church modes, with moderate positive agreement across diverse chord voicings. Across six paired seeds, DS also achieved the highest mean on every reported MusicQA and Music2Emo endpoint relative to the baseline, a parameter-matched Gaussian branch, and an architecture-matched magnitude-CQT branch. The Gaussian control indicates that capacity alone is insufficient, while the smaller advantage over magnitude CQT suggests value beyond an added pitch-resolved pathway, within the stated compression and normalization caveat. Together, these results support explicit relational structure as a useful and inspectable complement to learned music representations.

Beyond aggregate scores, the retained time--frequency layout gives DS a diagnostic role that scalar dissonance summaries cannot provide. Researchers can inspect when and where a modeled relation is concentrated, while the cross-reference construction can condition that attribution on an explicit tonic, chord, or spectral template. Because extraction is deterministic and cached, and the adapter preserves host shapes and offers exact fallback, the representation can be evaluated alongside existing systems without replacing their audio encoders. These properties make DS a concrete basis for theory-conditioned probing and model comparison, although their practical value outside the evaluated tasks remains to be tested.

The evidence does not establish DS as a complete model of consonance or musical preference. The kernel encodes one periodicity/harmonic-distance prior with deliberate octave folding and omits auditory-filter bandwidths, masking, tonal syntax, harsh high-frequency content, rhythmic tension, and culturally or individually learned preference. The controlled rankings are partly theory-derived, no new listening study was conducted, and automatic MusicQA references measure similarity rather than factual or expert-level harmonic reasoning. Evaluation is also limited to two host families and does not fully isolate the relation transform from branch-input compression and normalization. Future work should package the extraction and visualization pipeline as a reusable library, calibrate the representation with listener data and individual perceptual variation, and test its transfer to broader audio and symbolic tasks such as generation, style analysis, and standardized symbolic dissonance encoding.

\clearpage
\bibliography{references}
\end{document}


\maketitle
\enlargethispage{2pt}
\providecommand{\rmacro}{\overline{R^2}_{\mathrm{VA}}}

\section{Reproducibility Package}
The separately submitted Code and Data Archive contains the DS implementation, extraction configuration, cached-feature metadata schema, fixed dataset manifests, audio hashes, training configurations, environment lock files, checkpoint-selection rules, evaluation scripts, seed-level predictions, and scripts that regenerate every reported table. It includes scripts and instructions for obtaining public datasets and pretrained models from their cited official sources. The archive is packaged without author-identifying repository metadata.

\section{Full Derivation and Extensions of Dissonance Spectrum}
\label{sec:supp-derivation}
This section records the complete algebra and optional variants underlying the concise main-paper description and clarifies implementation shapes and experimental conventions.

\subsection{Preprocessing Conventions}
Let $\mathbf X(k,t)$ denote the magnitude CQT, $\vec{x}^{\,t}=\mathbf X(:,t)\in\mathbb R^K$, and $x_k^{\,t}=\mathbf X(k,t)$. The denoising and peak-selection operators are
\begin{equation}
\mathbf{X}_{\mathrm{denoised}}(k,t)
=
\mathbf{X}(k,t)
\cdot
\mathbf{1}
\left(
\mathbf{X}(k,t) \ge \theta
\right),
\end{equation}
\begin{equation}
\vec{x}^{\,t}_{\mathrm{peak}}
=
\vec{x}^{\,t}
\odot
\left[
\mathbf{1}
\left(
x_k^{\,t}
\text{ is a peak point in }
\vec{x}^{\,t}
\right)
\right]_{k=0}^{K-1}.
\end{equation}
Two normalization choices are useful in different settings:
\begin{equation}
\begin{aligned}
m(t)&=\max_k x_k^{\,t},
&\vec{x}^{\,t}_{\mathrm{frame}}&=\vec{x}^{\,t}/m(t),\\
m&=\max_{k,t}x_k^{\,t},
&\mathbf X_{\mathrm{global}}&=\mathbf X/m.
\end{aligned}
\end{equation}
The downstream model comparisons use excerpt-level global-maximum normalization $m$, matching the magnitude-CQT control and avoiding framewise loudness equalization. The controlled music-theory tests instead use the context-window normalization described below, with target and reference spectra normalized separately. Each comparison uses one fixed convention recorded in its configuration.

\subsection{Direct Pairwise Form}
The intrinsic element is
\begin{equation}
d^t_k =
\frac{1}{K}
\sum_{l=0}^{K-1}
x^t_k\, x^t_l\, \bar{D}(p_l-p_k),
\end{equation}
and the complete representation is
\begin{equation}
\mathbf D=
\begin{bmatrix}
\vec d^{\,0} & \vec d^{\,1} & \cdots & \vec d^{\,T-1}
\end{bmatrix}
\in \mathbb R^{K\times T}.
\end{equation}
The factor $1/K$ averages the reference-bin contributions and removes their direct linear count factor under otherwise matched grids; it is not a guarantee of invariance to CQT resolution. The direct pair-attribution concept is visualized in the main paper.

\subsection{Index Derivation of the Correlation Form}
\begin{figure}[H]
\centering
\includegraphics[width=.72\columnwidth]{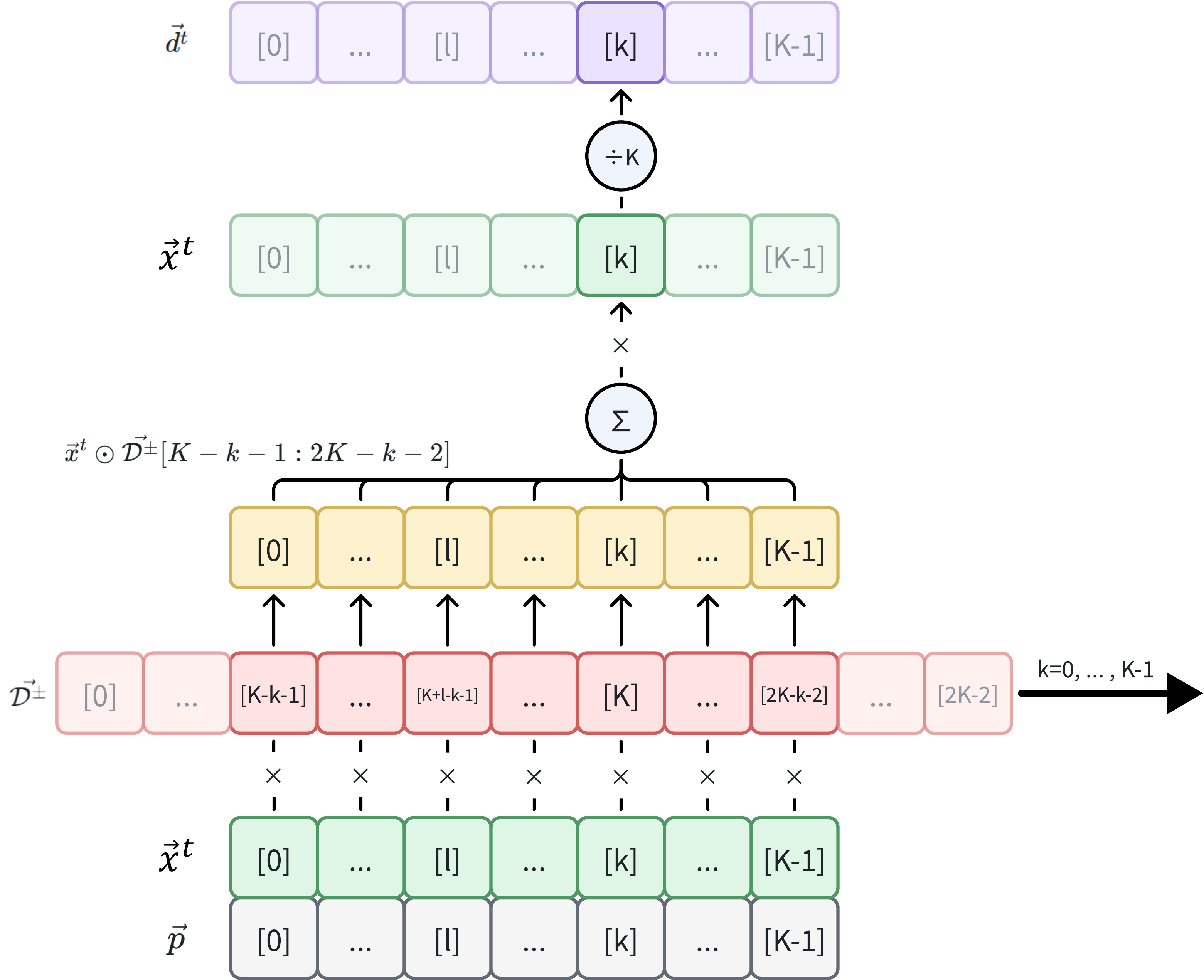}
\caption{Index alignment for the frequency-axis correlation form of intrinsic DS.}
\label{fig:supp-corr}
\end{figure}

The CQT pitch grid and the zero-indexed stored relation vector are
\begin{equation}
p_k=\mathrm{pitch}_{\min}+\frac{12k}{B},
\qquad k=0,\ldots,K-1,
\end{equation}
\begin{equation}
\left[\vec{\mathcal D^\pm}\right]_j
=
\bar D\!\left(\frac{12(j-K+1)}{B}\right),
\qquad j=0,\ldots,2K-2.
\end{equation}
Because the kernel depends only on pitch difference,
\begin{equation}
\bar{D}(p_l-p_k)
=
\vec{\mathcal{D}^{\pm}}_{K+l-k-1}.
\end{equation}
For a length-$(2K-1)$ vector $\vec a$ and a length-$K$ vector $\vec b$, the implementation uses the target-aligned convention
\begin{equation}
[\vec a\star_{\mathrm{valid},f}\vec b]_k
=
\sum_{l=0}^{K-1}a_{K+l-k-1}b_l,
\quad k=0,\ldots,K-1.
\label{eq:supp-aligned-correlation}
\end{equation}
With increasing-lag storage, this equals a standard valid cross-correlation followed by a fixed frequency-axis reversal; an equivalent lag-reversed layout avoids the explicit reversal. Consequently,
\begin{align}
d^{\,t}_k
&=
\frac{1}{K}x_k^t
\sum_{l=0}^{K-1}
x^t_l\,\bar{D}(p_l-p_k)\\
&=
\frac{1}{K}x_k^t
\sum_{l=0}^{K-1}
x_l^t\,\mathcal D^\pm_{K+l-k-1}.
\end{align}
Sliding the complete relation vector over the CQT gives the reference-relation frame
\begin{equation}
\vec{r}^{\,t}=\vec{\mathcal{D}^\pm}
\star_{\mathrm{valid},f}
\vec{x}^{\,t},
\end{equation}
and therefore
\begin{equation}
\vec{d}^{\,t}
=
\frac{1}{K}
\vec{r}^{\,t}
\odot
\vec{x}^t
=
\frac{1}{K}
\left(
\vec{\mathcal{D}^\pm}
\star_{\mathrm{valid},f}
\vec{x}^{\,t}
\right)
\odot
\vec{x}^t.
\end{equation}
Figure~\ref{fig:supp-corr} visualizes the index alignment. Equation~\ref{eq:supp-aligned-correlation} is applied independently at every time frame and returns the $K$ target locations in their original frequency order.

\begin{figure}[H]
\centering
\includegraphics[width=.72\columnwidth]{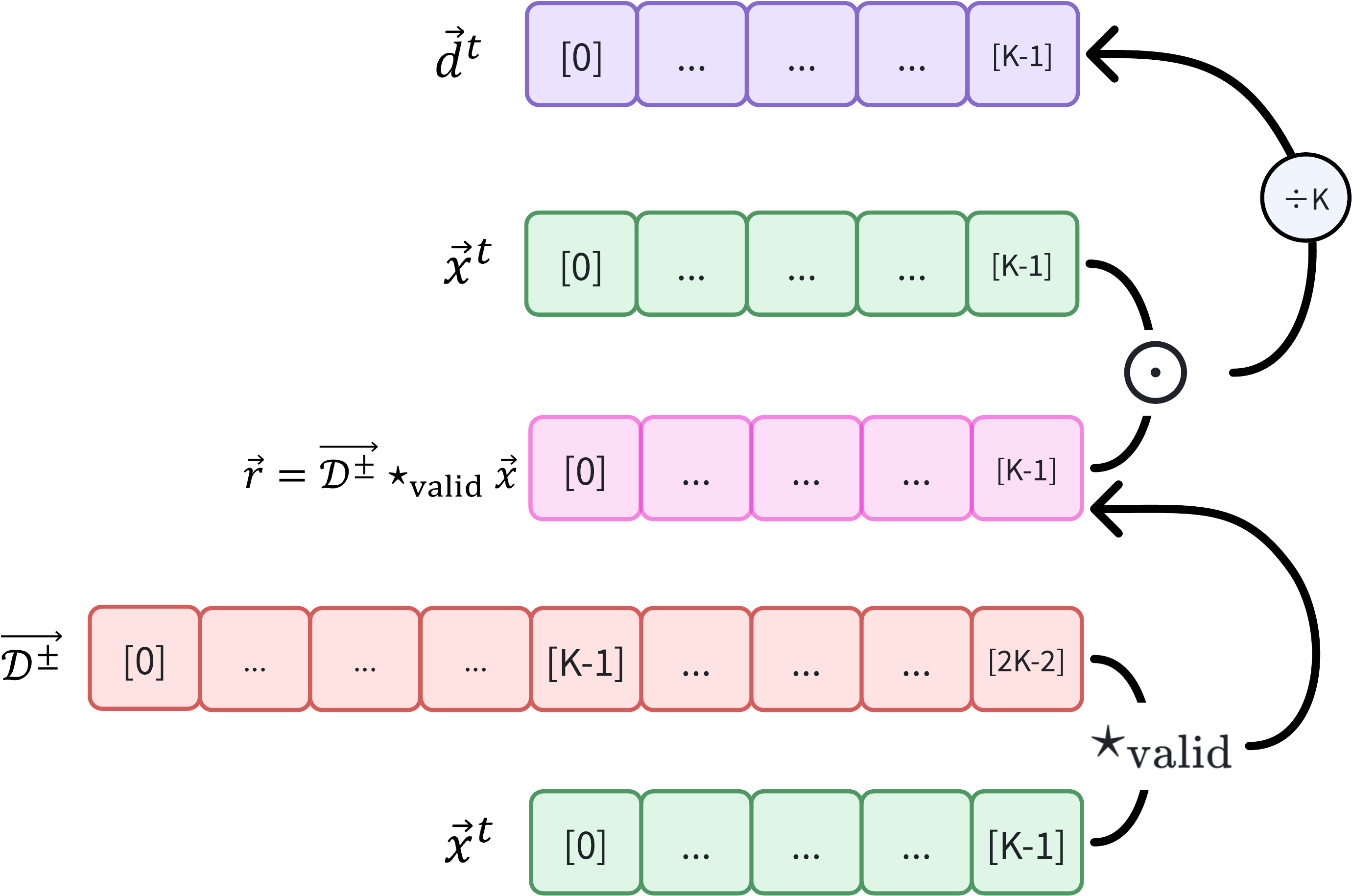}
\caption{Vectorized intrinsic-DS computation after replacing the explicit pairwise sum by valid correlation.}
\label{fig:supp-intrinsic-simplify}
\end{figure}

\subsection{Cross Dissonance Spectrum}
\begin{figure}[H]
\centering
\includegraphics[width=.72\columnwidth]{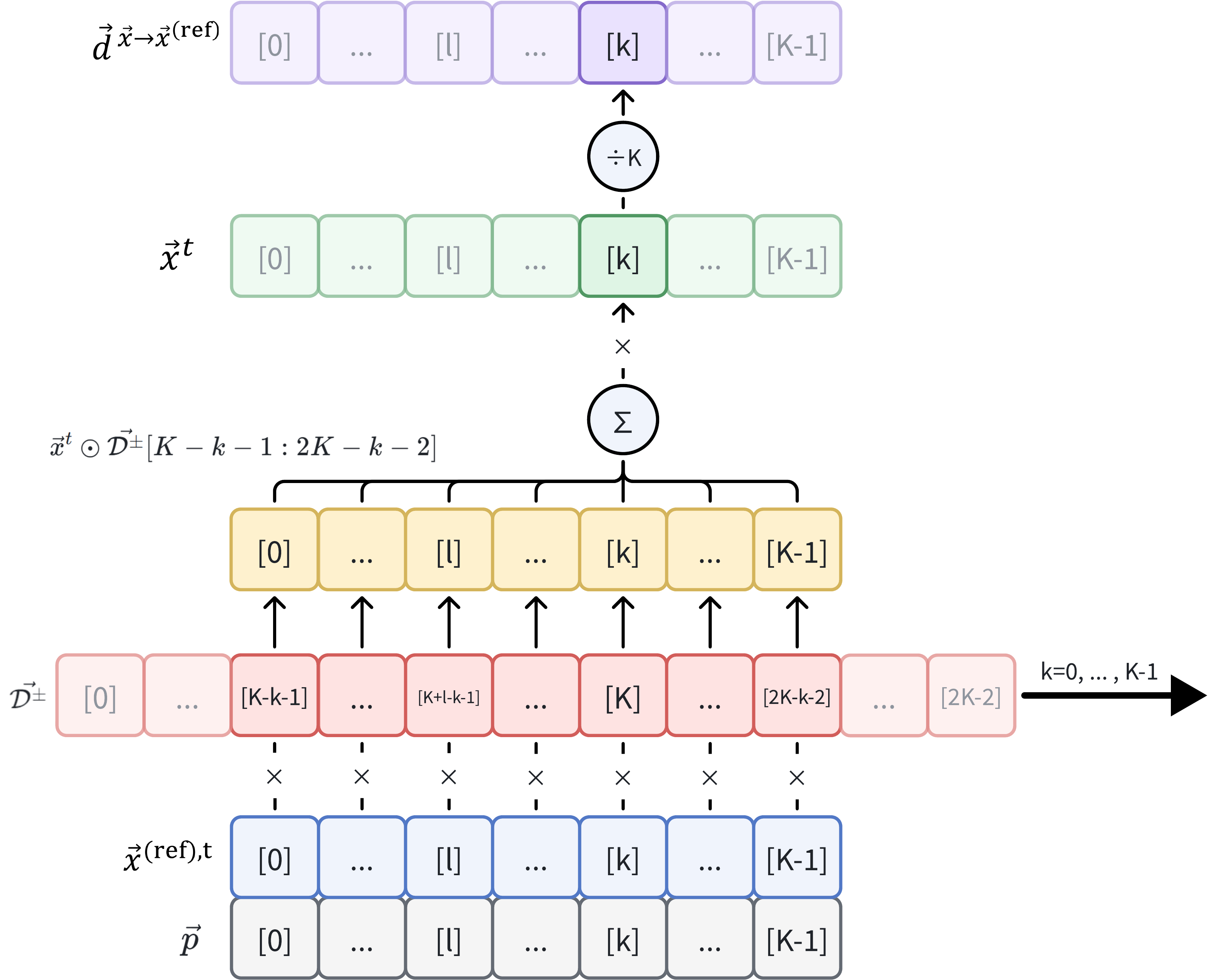}
\caption{Index alignment for cross-DS correlation using a fixed or time-varying reference spectrum.}
\label{fig:supp-cross}
\end{figure}

For target $\vec x$ and an arbitrary reference spectrum $\vec{x}^{(\mathrm{ref})}$, the cross-spectrum is
{\small
\begin{align}
d_{k}^{\,\vec{x} \to \vec{x}^{(\mathrm{ref})}}
&=
\frac{1}{K}
\sum_{l=0}^{K-1}
x_{k}x_{l}^{(\mathrm{ref})}\,
\bar{D}(p_l-p_k)\\
&=
\frac{1}{K}
x_k
\sum_{l=0}^{K-1}
x_l^{(\mathrm{ref})}\,
\mathcal D^\pm_{K+l-k-1}.
\end{align}
}
Its reusable reference map and final attribution are
\begin{equation}
\vec{r}=\vec{\mathcal{D}^\pm}
\star_{\mathrm{valid},f}
\vec{x}^{(\mathrm{ref})},
\end{equation}
\begin{equation}
\vec{d}^{\,\vec{x} \to \vec{x}^{(\mathrm{ref})}}
=
\frac{1}{K}
\left(
\vec{\mathcal{D}^\pm}
\star_{\mathrm{valid},f}
\vec{x}^{(\mathrm{ref})}
\right)
\odot\vec{x}.
\end{equation}
Intrinsic DS is the special case $\vec{x}^{(\mathrm{ref})}=\vec x$. A fixed reference may instead encode a tonic template or an instrument tone. Figure~\ref{fig:supp-cross} shows the corresponding cross-reference computation.

\begin{figure}[H]
\centering
\includegraphics[width=.72\columnwidth]{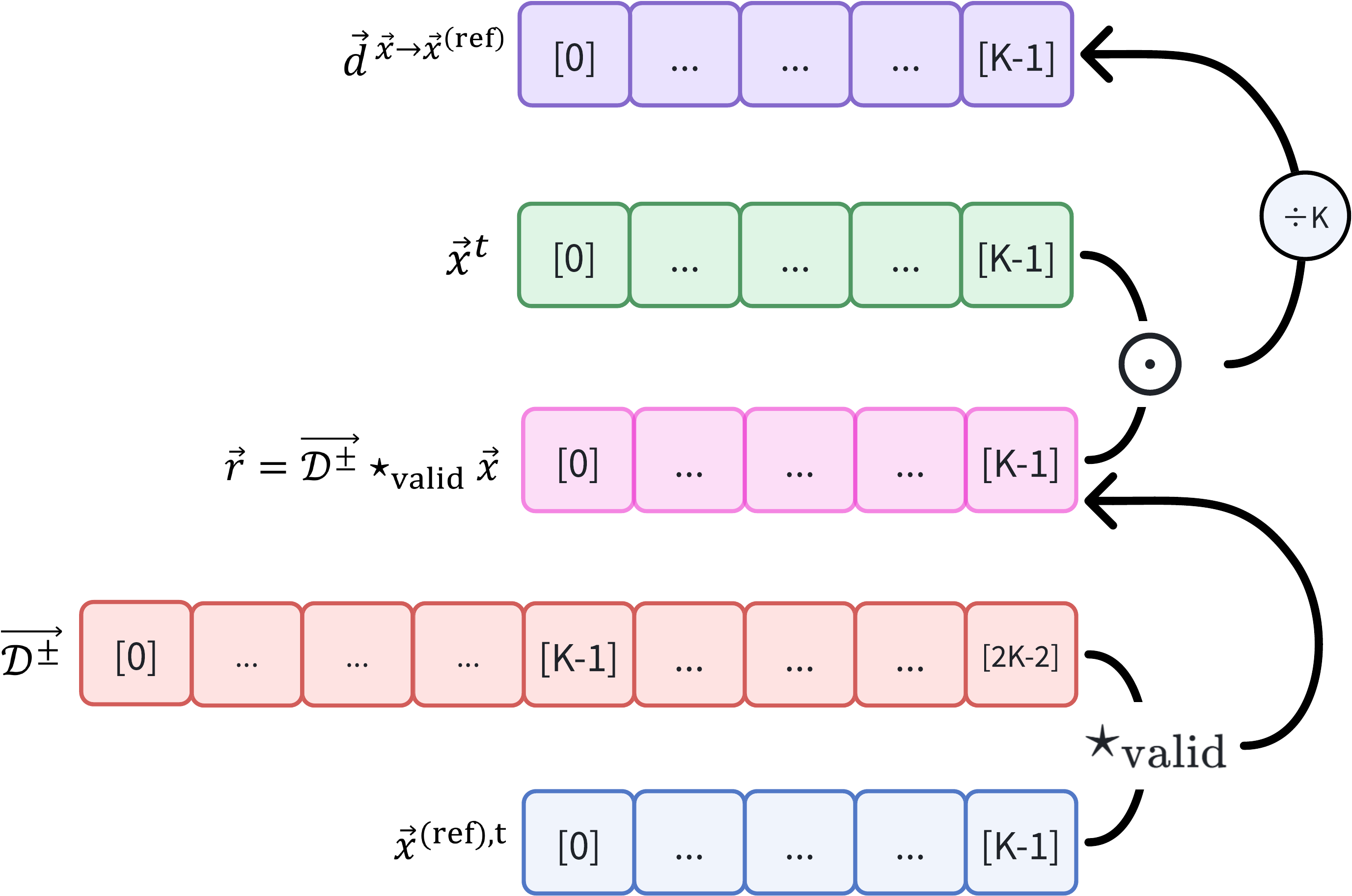}
\caption{Vectorized cross-DS computation using a reusable reference-relation vector.}
\label{fig:supp-cross-simplify}
\end{figure}

\subsection{Matrix and Temporal Variants}
Repeating the same relation vector across time gives
\begin{equation}
\mathbf{\mathcal{D}}^{\pm}
=
\begin{bmatrix}
\vec{\mathcal{D}}^{\pm}&\vec{\mathcal{D}}^{\pm}&\cdots&\vec{\mathcal{D}}^{\pm}
\end{bmatrix}
\in\mathbb{R}^{(2K-1)\times T}.
\end{equation}
For a general reference CQT,
\begin{equation}
\mathbf{R}
=
\mathbf{\mathcal{D}}^{\pm}
\star_{\mathrm{valid},f}
\mathbf{X}^{(\mathrm{ref})},
\end{equation}
\begin{equation}
\mathbf{D}
=\frac{1}{K}\,\mathbf{R}\odot\mathbf{X}
=\frac{1}{K}\,
(\mathbf{\mathcal{D}}^{\pm}
\star_{\mathrm{valid},f}
\mathbf{X}^{(\mathrm{ref})})\odot\mathbf{X}.
\end{equation}
Both $\mathbf R$ and $\mathbf D$ have shape $K\times T$ under the frequency-axis valid-correlation convention used here. Optional temporal DS averages references from the preceding $n$ frames:
\begin{equation}
\vec{d}^{\,(\mathrm{temporal}),\,t}
=
\frac{1}{n}
\sum_{i=0}^{n-1}
\vec{d}^{\left(
\vec{x}^{\,t}
\to
\vec{x}^{\,t-i}
\right)}
=
\vec{d}^{\left(
\vec{x}^{\,t}
\to
\frac{1}{n}
\sum_{i=0}^{n-1}
\vec{x}^{\,t-i}
\right)}.
\end{equation}
A tonic--intrinsic combination is
\begin{equation}
\vec{d}^{\,(\mathrm{combine}),\,t}
=
w_{\mathrm{intrinsic}}\vec{d}^{\,t}
+
w_{\mathrm{tonic}}\vec{d}^{\,(\mathrm{tonic}),\,t}.
\end{equation}
The downstream model comparisons use intrinsic DS. The controlled music-theory tests use tonic, chord, or tonic--intrinsic references to isolate the relation being examined.

\section{Controlled Music-Theory Validation Details}
\label{sec:supp-theory}

\subsection{Stimuli, References, and Aggregation}
The validation suite uses MIDI-rendered stimuli with fixed note events and piano timbre unless otherwise stated. Nominal durations describe MIDI events; WAV files include onset and release tails. The controlled tests are not a listener study. They ask whether the relation operator and its audio realization reproduce predefined ordinal structures under transparent reference choices.

\begin{table}[H]
\centering
\scriptsize
\caption{Controlled-validation configuration.}
\label{tab:supp-theory-config}
\resizebox{\columnwidth}{!}{
\begin{tabular}{ll}
\toprule
Item & Setting \\
\midrule
Audio / frame rate & 22.05 kHz mono / 4 frames s$^{-1}$ \\
CQT grid & 8 octaves, 72 bins octave$^{-1}$, $K=576$, $f_{\min}=\mathrm{C1}$ \\
Kernel & $Q=60$, relative tolerance $\alpha=.01$, octave folding \\
Preprocessing & linear magnitude, relative-frame soft gate at 0.1, context window 4 s \\
Peak selection & prominence .015; other peak filters disabled \\
Scaling & target and reference normalized separately; final factor $1/K$ \\
Intervals / scales & C4 tonic reference \\
Chord qualities & equal-weight C4-tonic and intrinsic DS \\
Chord connections & C-major chord reference \\
Aggregation & $D_{\max}$ for intervals/chords/connections; $D_{\mathrm{sum}}$ for scales \\
\bottomrule
\end{tabular}}
\end{table}

For a frame, $D(t)=\sum_k\mathbf D(k,t)$. We use $D_{\max}=\max_t D(t)$ when a stimulus contains a sustained interval or chord event and $D_{\mathrm{sum}}=\sum_tD(t)$ when an entire sequential scale is the analysis unit. Rank hypotheses are min--max mapped only for visualization; all reported correlations use the original ranks and unscaled DS summaries.

\begin{figure}[H]
\centering
\includegraphics[width=.96\columnwidth]{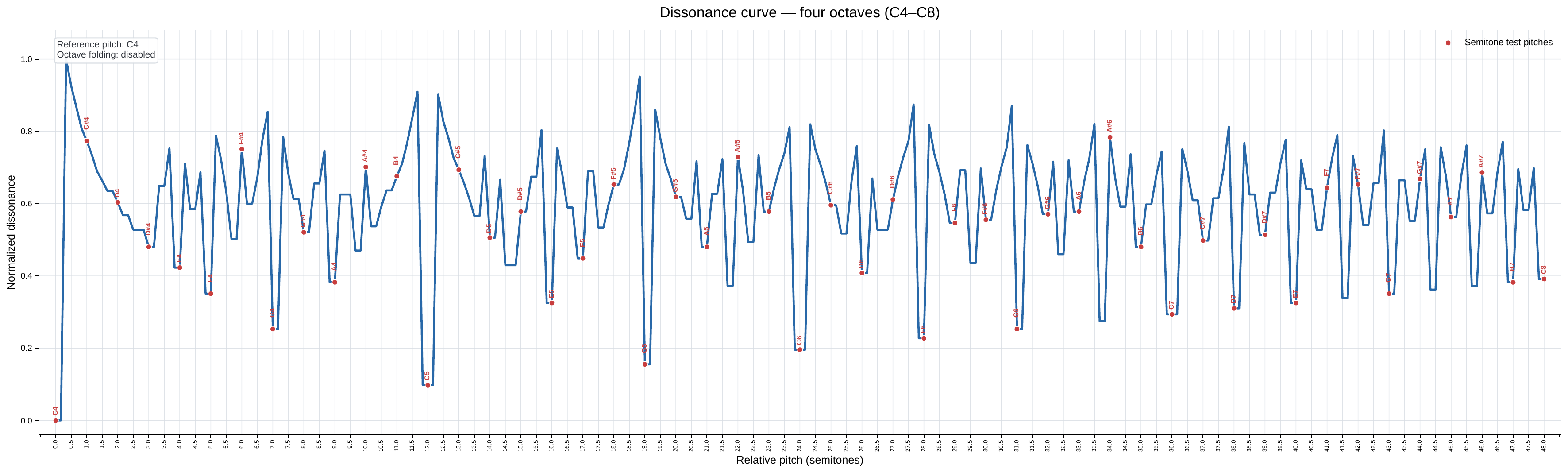}
\caption{Four-octave view of the normalized relation kernel, illustrating its octave-folded repetition. Integer-semitone samples are marked for reference.}
\label{fig:supp-kernel}
\end{figure}

\subsection{Intervals}
Thirteen notes from C4 to C5 are compared against a C4 reference. The historical order used by the validation assets is related to classic dyad-consonance experiments \citep{malmberg1918consonance,schwartz2003statistical}, but it is not presented as a direct transcription of any single published table.

\begin{table}[H]
\centering
\scriptsize
\caption{Interval results under the C4 tonic reference.}
\label{tab:supp-intervals}
\begin{tabular}{lrr@{\qquad}lrr}
\toprule
Note & Rank & $D_{\max}$ & Note & Rank & $D_{\max}$ \\
\midrule
C4 & 1 & .000253 & G4 & 3 & .001145 \\
C$\sharp$4 & 13 & .004379 & G$\sharp$4 & 7 & .001913 \\
D4 & 11 & .003476 & A4 & 6 & .001703 \\
D$\sharp$4 & 8 & .002463 & A$\sharp$4 & 10 & .002583 \\
E4 & 5 & .002298 & B4 & 12 & .002802 \\
F4 & 4 & .001614 & C5 & 2 & .000030 \\
F$\sharp$4 & 9 & .003391 & & & \\
\bottomrule
\end{tabular}
\end{table}

Spearman $\rho=.95055$ ($p=6.36\times10^{-7}$) and Kendall $\tau_b=.84615$ ($p=5.20\times10^{-6}$). The minor second is maximal, the tritone is high, the perfect fifth is low, and the octave is minimal. The small nonzero unison value reflects slight differences between independently rendered target and reference recordings.

\begin{figure}[H]
\centering
\includegraphics[width=\columnwidth]{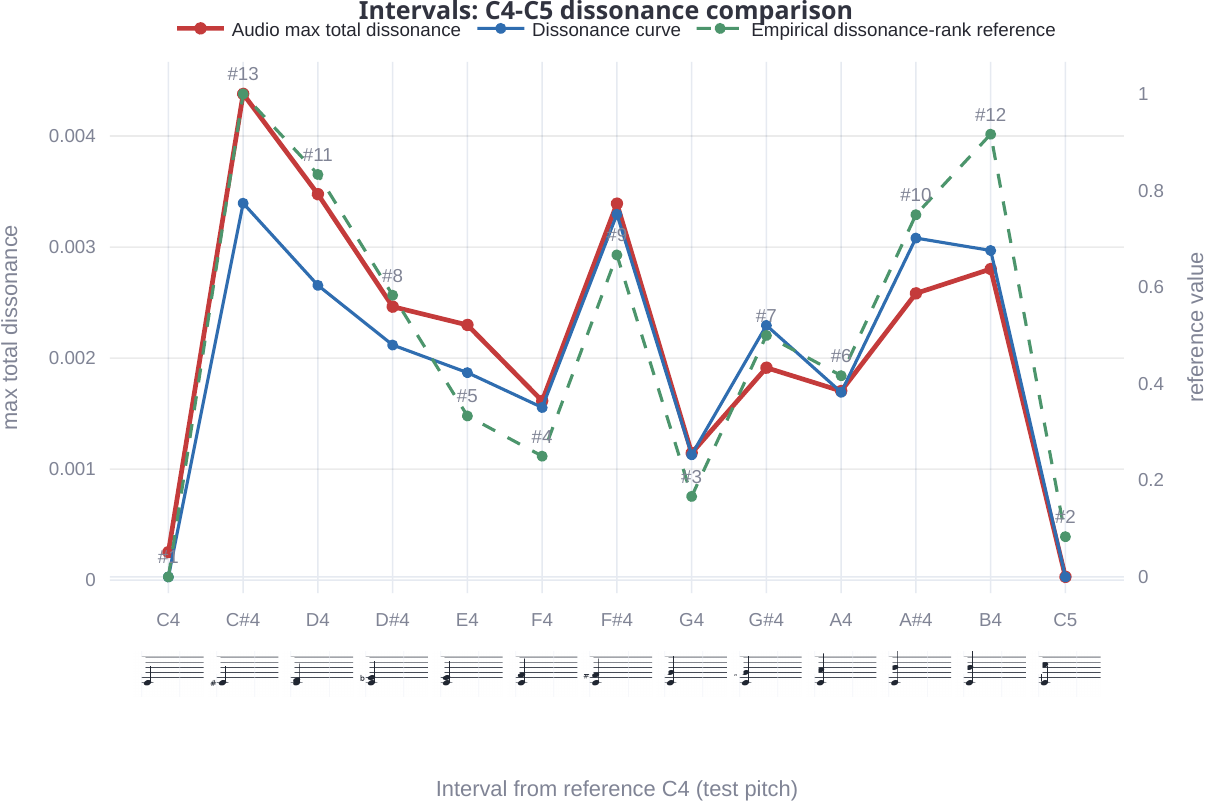}
\caption{Interval validation details: audio DS maxima, sampled kernel values, and the predefined rank reference.}
\label{fig:supp-interval-summary}
\end{figure}

\begin{figure}[H]
\centering
\includegraphics[width=\columnwidth]{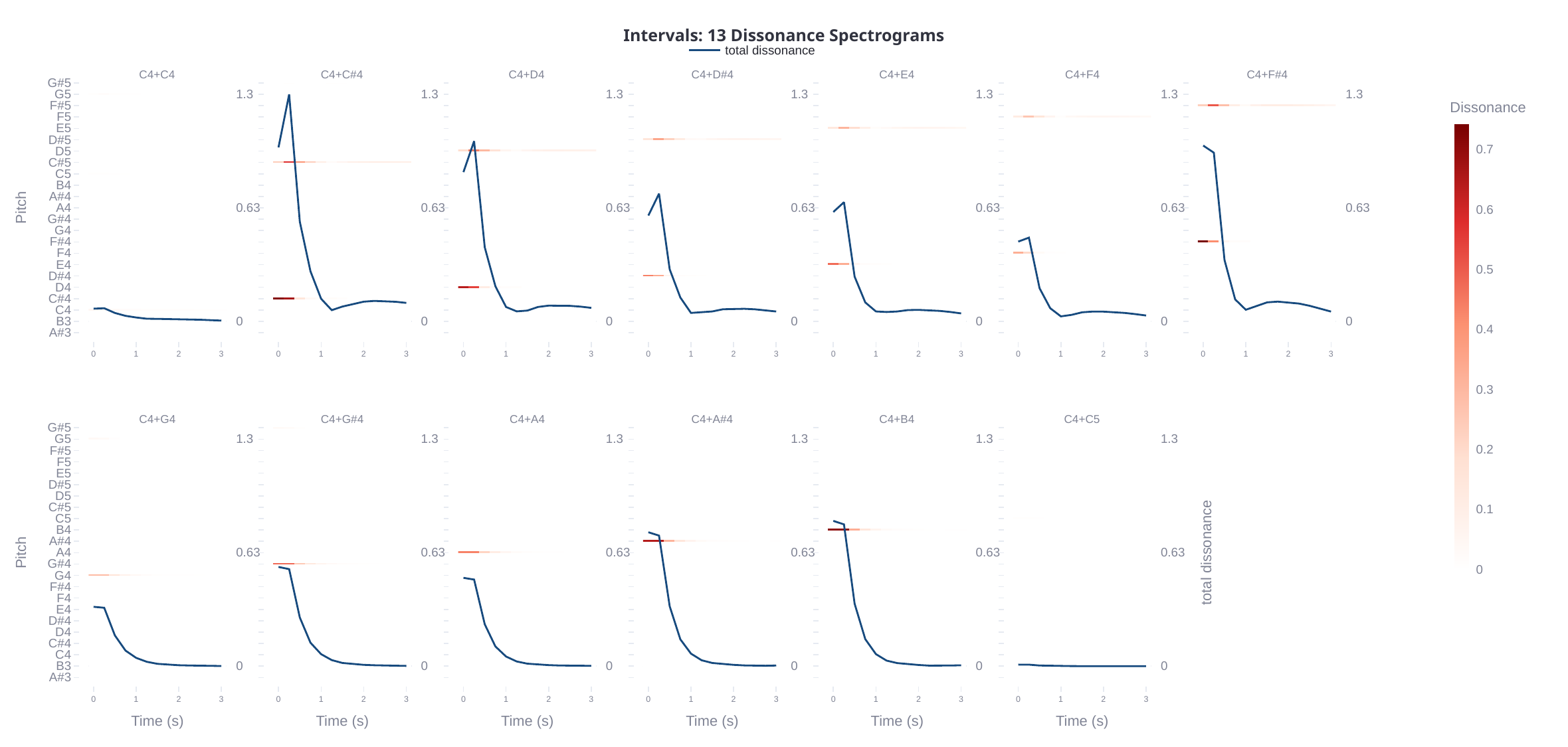}
\caption{Representative interval DS maps for unison, minor second, perfect fifth, and octave.}
\label{fig:supp-interval-maps}
\end{figure}

\subsection{Chord Qualities}
The four representative classes are perfectly ordered as major $<$ minor $<$ suspended $<$ diminished. Because $n=4$, this is treated as an ordering check rather than a meaningful significance test. The larger set varies chord quality and voicing and therefore provides a stricter test.

\begin{table*}[t]
\centering
\scriptsize
\caption{Extended chord-quality validation. Semitone arrays are relative to C4.}
\label{tab:supp-chords}
\resizebox{\textwidth}{!}{
\begin{tabular}{llrr@{\qquad}llrr}
\toprule
Chord & Semitones & Rank & $D_{\max}$ & Chord & Semitones & Rank & $D_{\max}$ \\
\midrule
Major-1 & [0,4,7] & 1 & .005691 & Sus-1 & [0,7,17] & 7 & .005641 \\
Major-2 & [0,3,8] & 5 & .006941 & Sus-2 & [0,2,7] & 6 & .006758 \\
Major-3 & [0,9,17] & 3 & .006299 & Sus-3 & [0,10,17] & 4 & .007769 \\
Minor-1 & [0,3,7] & 2 & .006058 & Dim-1 & [0,3,6] & 12 & .015950 \\
Minor-2 & [0,4,9] & 10 & .006738 & Dim-2 & [0,9,15] & 9 & .008581 \\
Minor-3 & [0,8,17] & 8 & .006363 & Dim-3 & [0,9,18] & 11 & .008172 \\
Aug & [0,4,8] & 13 & .007305 & & & & \\
\bottomrule
\end{tabular}}
\end{table*}

Across 13 voicings, Spearman $\rho=.62637$ ($p=.02199$) and Kendall $\tau_b=.46154$ ($p=.03048$). The positive association coexists with local reversals, such as Sus-1 below Major-1 and the augmented triad below Dim-1. These deviations are informative rather than errors: DS analyzes realized spectra and therefore responds to spacing and inversion instead of assigning one constant to a chord label.

\begin{figure}[H]
\centering
\includegraphics[width=\columnwidth]{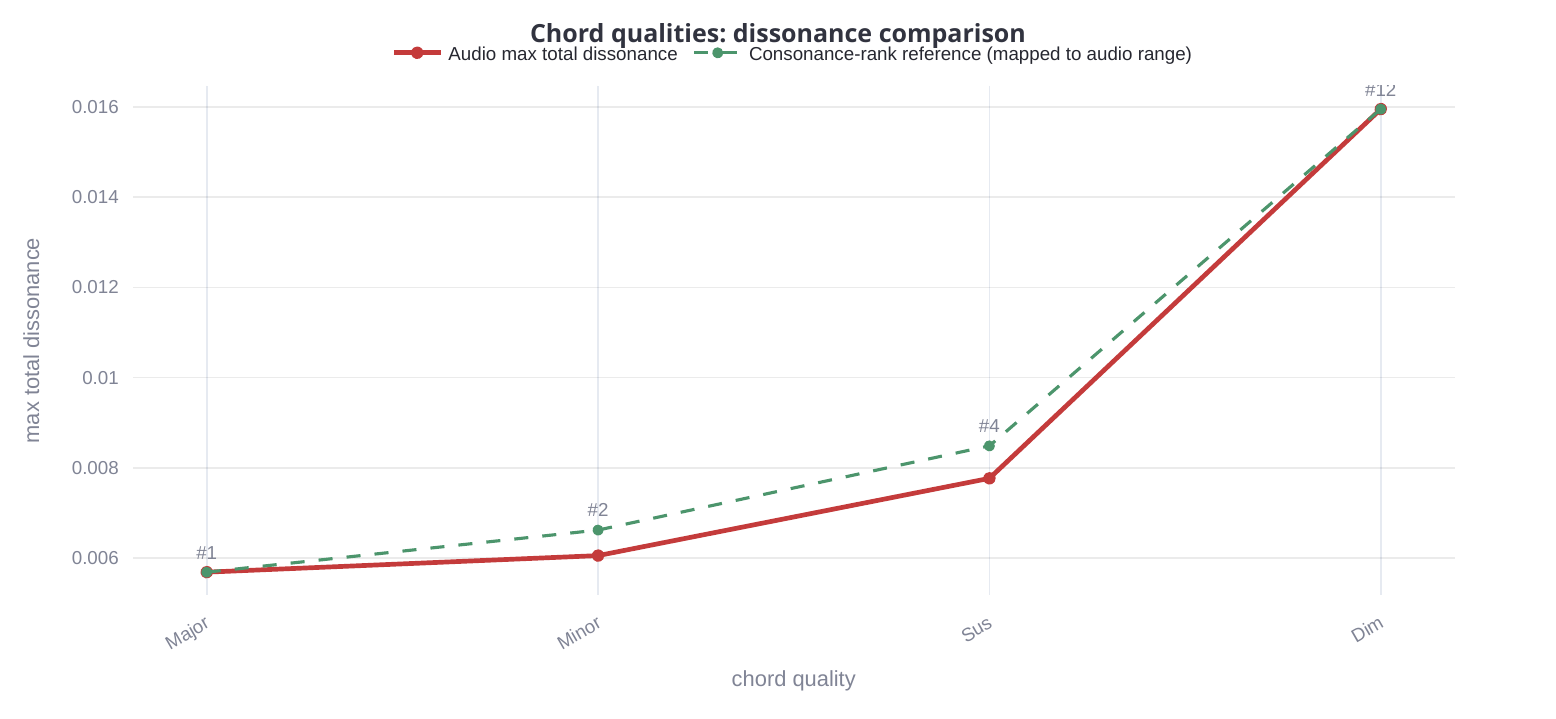}
\caption{Representative chord-quality ordering.}
\label{fig:supp-chord-selected}
\end{figure}

\begin{figure}[H]
\centering
\includegraphics[width=\columnwidth]{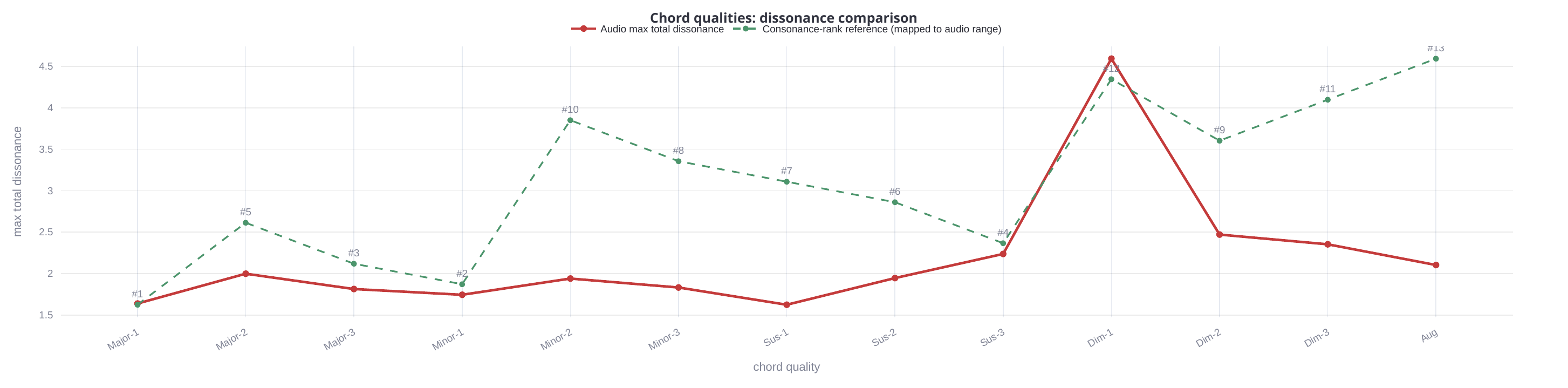}
\caption{Extended chord-quality and voicing results.}
\label{fig:supp-chord-full}
\end{figure}

\begin{figure}[H]
\centering
\includegraphics[width=\columnwidth]{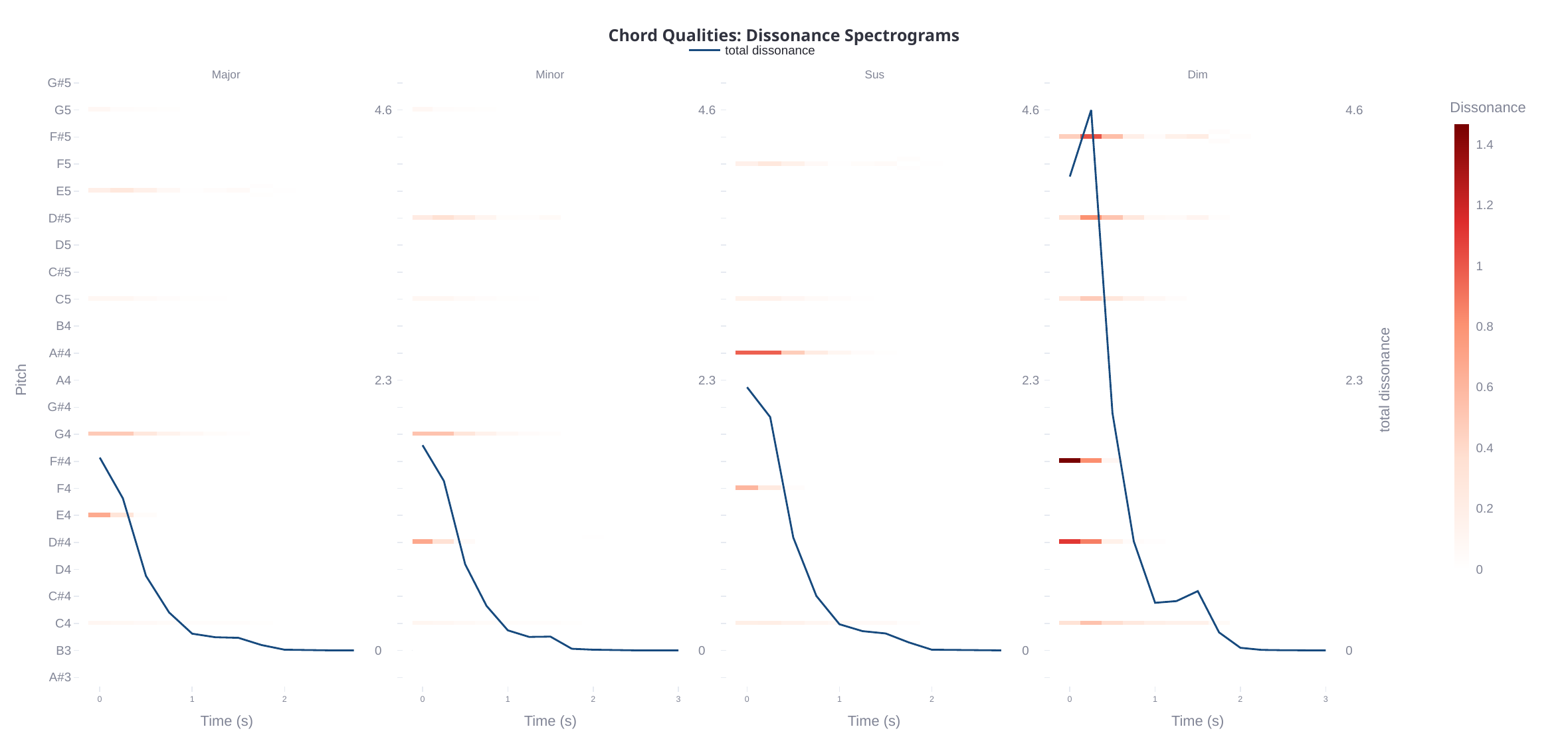}
\caption{Representative tonic-referenced chord-quality spectra.}
\label{fig:supp-chord-spectra}
\end{figure}

\subsection{Functional Chord Connections}
Each stimulus plays one diatonic chord followed by C major. The ordinal code assigns 1 to tonic-function chords (I, iii, vi), 2 to predominant chords (ii, IV), and 3 to dominant-function chords (V7, vii$^\circ$). This reference-conditioned construction operationalizes a connection as the first chord's relation to C major; it does not model voice leading or learned temporal expectation. The code is a theory-derived trend hypothesis, not a psychophysical scale.

\begin{table}[H]
\centering
\scriptsize
\caption{Chord connections to C major.}
\label{tab:supp-connections}
\begin{tabular}{lclrr}
\toprule
File & First chord & Group & Rank & $D_{\max}$ \\
\midrule
CC & C--E--G & tonic & 1 & .003848 \\
DC & D--F--A & predominant & 2 & .009701 \\
EC & E--G--B & tonic & 1 & .005311 \\
FC & F--A--C & predominant & 2 & .008495 \\
G7C & G--B--D--F & dominant & 3 & .010312 \\
AC & A--C--E & tonic & 1 & .005695 \\
BC & B--D--F & dominant & 3 & .006699 \\
\bottomrule
\end{tabular}
\end{table}

Spearman $\rho=.79373$ ($p=.03310$) and Kendall $\tau_b=.65465$ ($p=.05363$). The overall group trend is recovered, although vii$^\circ$ and V7 differ substantially, showing that group membership does not determine the complete spectral value.

\begin{figure}[H]
\centering
\includegraphics[width=\columnwidth]{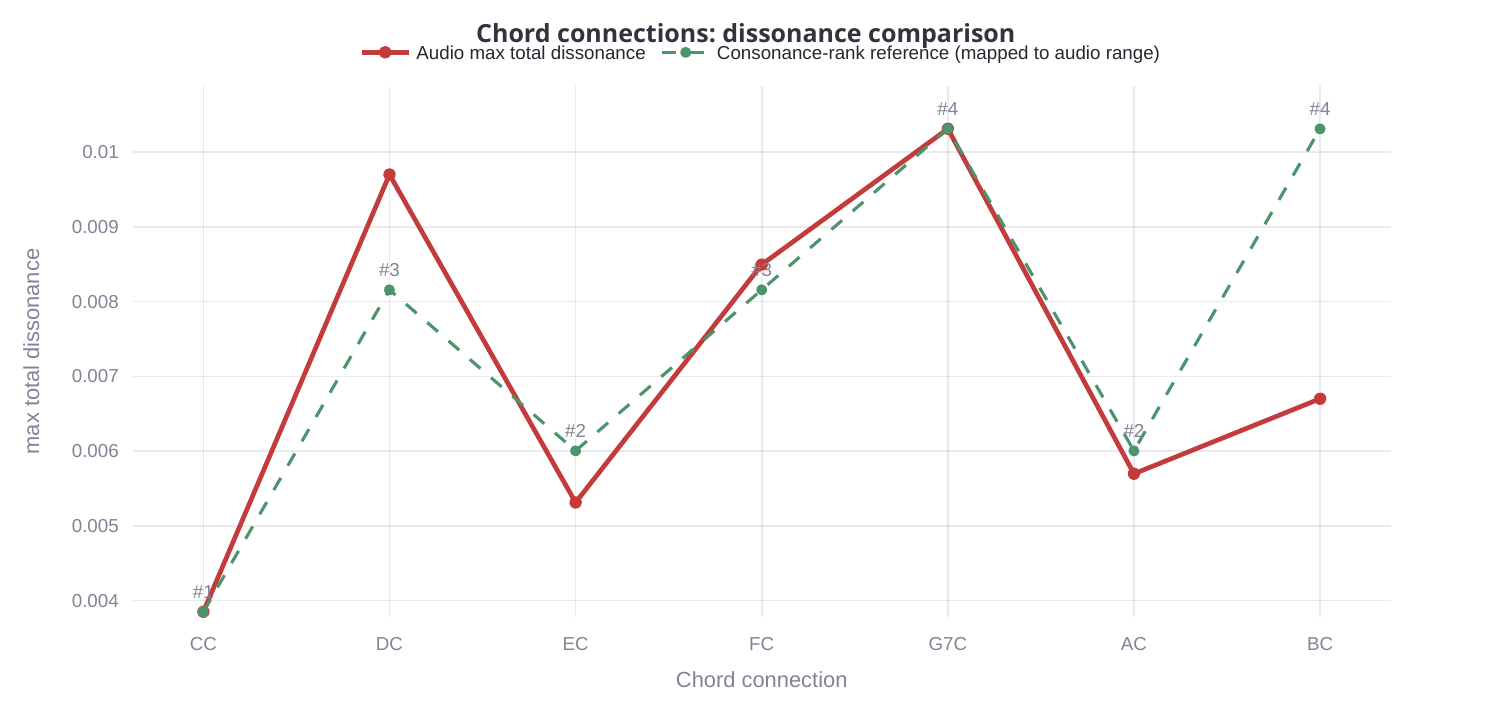}
\caption{Functional chord-connection results.}
\label{fig:supp-connections}
\end{figure}

\begin{figure}[H]
\centering
\includegraphics[width=\columnwidth]{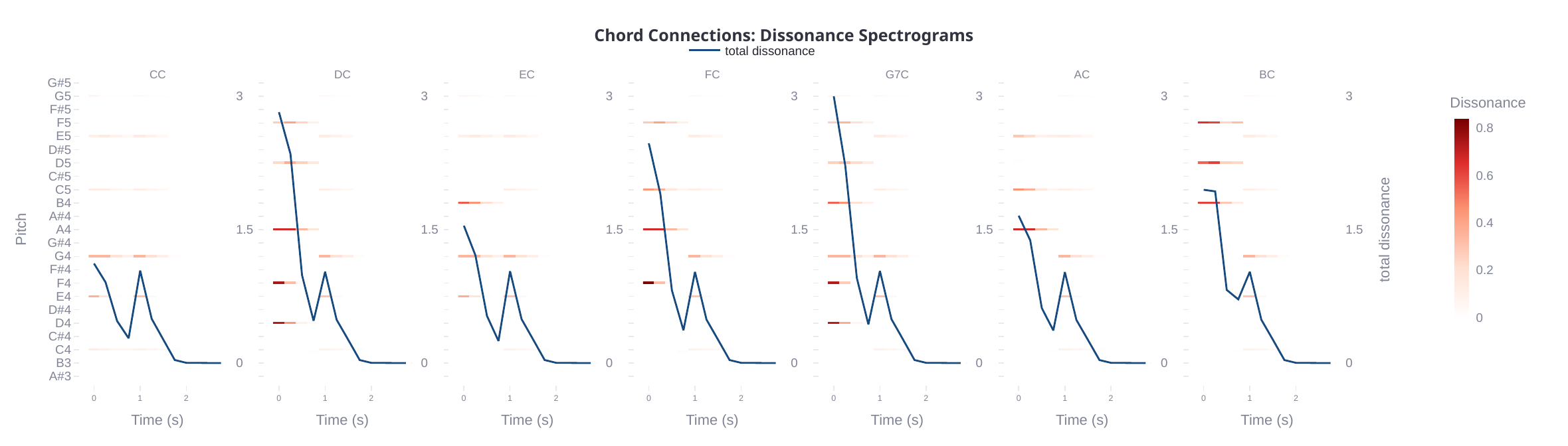}
\caption{Reference-conditioned spectra for the seven chord connections.}
\label{fig:supp-connection-spectra}
\end{figure}

\subsection{Scales and Modes}
Scale stimuli ascend from C4 with one second per note. For the seven church modes, $D_{\mathrm{sum}}$ correlates with the predefined order at Spearman $\rho=.89286$ ($p=.00681$) and Kendall $\tau_b=.80952$ ($p=.0107$).

\begin{table}[H]
\centering
\scriptsize
\caption{Church-mode results.}
\label{tab:supp-modes}
\begin{tabular}{lrr}
\toprule
Mode & Rank & $D_{\mathrm{sum}}$ \\
\midrule
Ionian & 1 & .030220 \\
Mixolydian & 2 & .030273 \\
Lydian & 3 & .033697 \\
Dorian & 4 & .030704 \\
Aeolian & 5 & .031444 \\
Phrygian & 6 & .033937 \\
Locrian & 7 & .038779 \\
\bottomrule
\end{tabular}
\end{table}

\begin{figure}[H]
\centering
\includegraphics[width=.82\columnwidth]{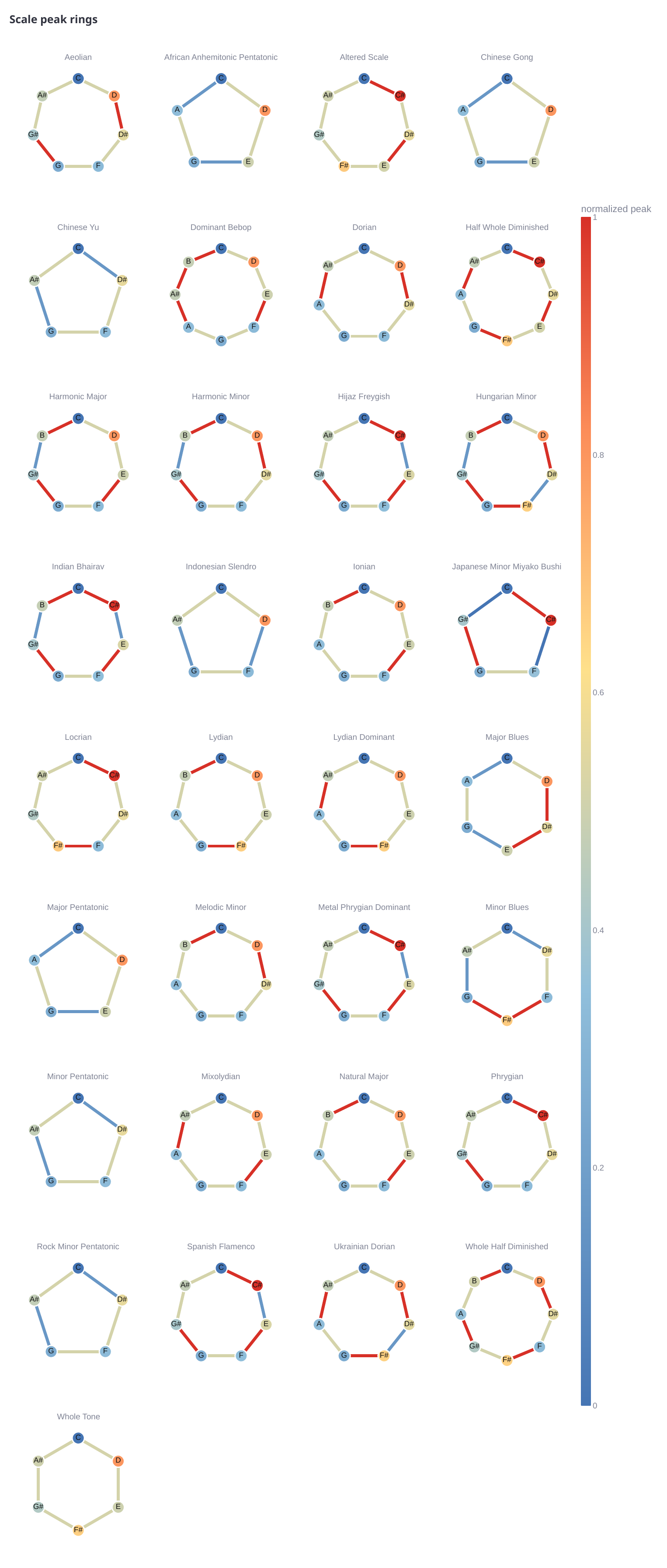}
\caption{Circular peak patterns for all available scale recordings. Only the seven church modes are used for the ordinal correlation in the main paper.}
\label{fig:supp-all-scales}
\end{figure}

The trend is strong but not strictly monotonic because Dorian lies below Lydian. This test aggregates tonic-relative interval content and should not be read as a complete model of modal perception. Figure~\ref{fig:supp-all-scales} includes all 33 available scale files. For scales without an externally grounded rank, the values should be interpreted as a quantitative ``consonance palette'': DS can compare and visualize their realized interval content, but the ordering is not claimed as a universal preference scale. The code defines a natural-minor file that is absent; the equivalent Aeolian pitch-class set is available.

\begin{figure}[H]
\centering
\includegraphics[width=.94\columnwidth]{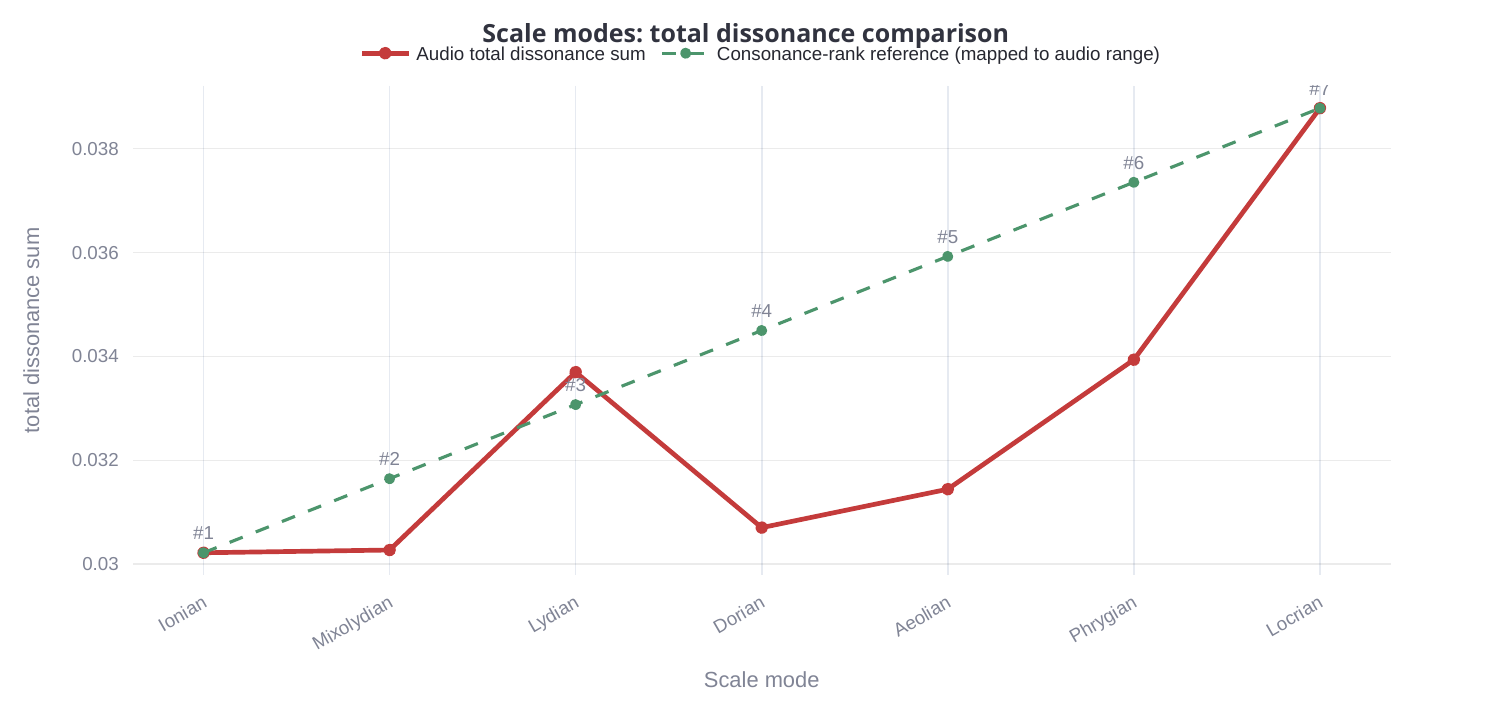}
\caption{Church-mode total DS and the predefined ordinal reference used in the main-paper correlation.}
\label{fig:supp-mode-summary}
\end{figure}

\begin{figure}[H]
\centering
\includegraphics[width=\columnwidth]{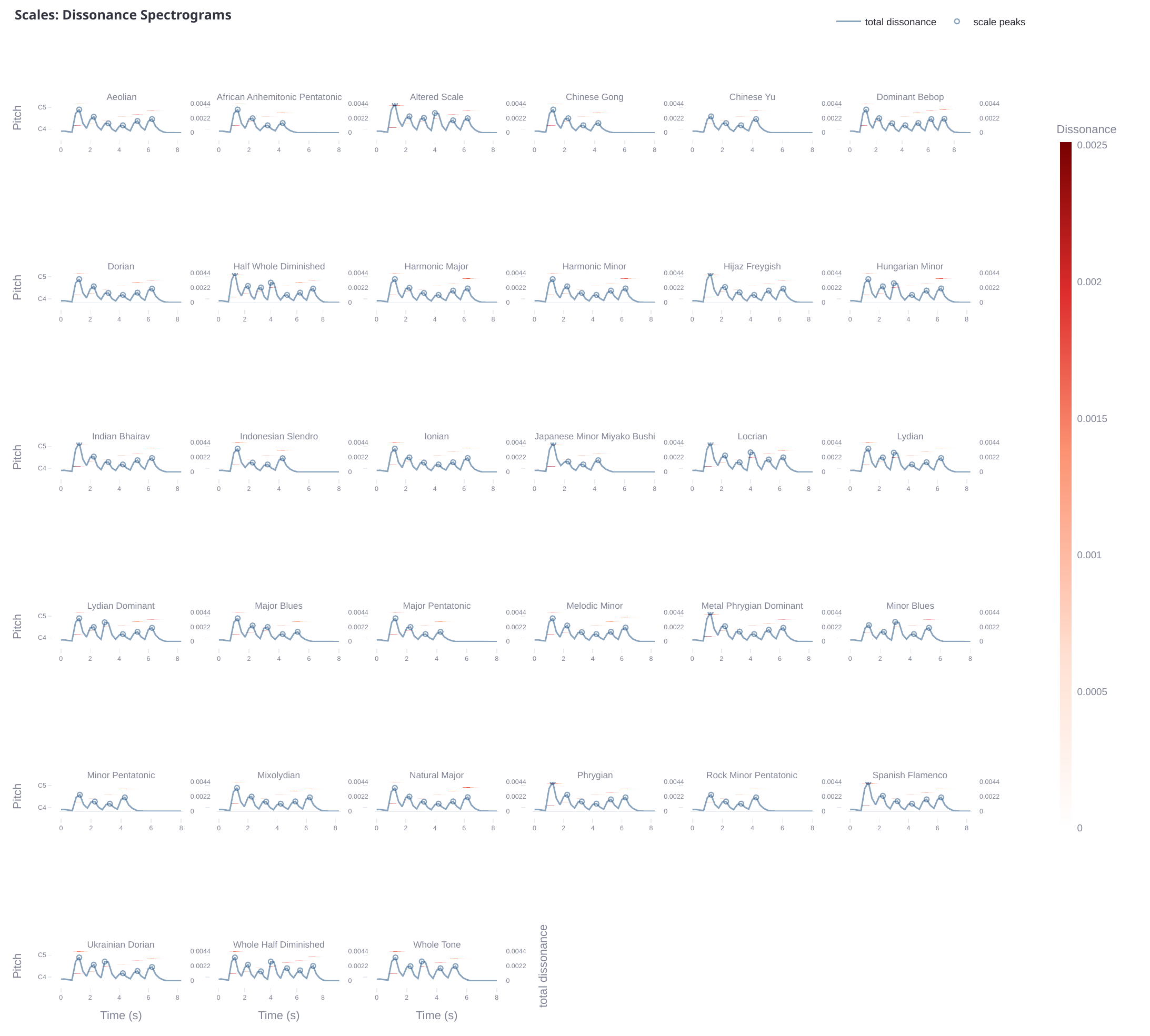}
\caption{DS curves for the available scale recordings, illustrating scale-dependent consonance profiles.}
\label{fig:supp-scale-spectra}
\end{figure}

\subsection{Timbre and Microtonal Examples}
The same C4--C5 chromatic sequence yields different maxima across seven software instruments, demonstrating sensitivity to partial amplitudes, envelopes, and noise. Because loudness, spectral centroid, and envelope are not matched, these values are descriptive comparisons of complete rendered sounds rather than a causal timbre ranking. A 24-tone-equal-temperament sequence further demonstrates that the continuous kernel can analyze 50-cent steps; no experiential ranking is imposed.

\begin{table}[H]
\centering
\scriptsize
\caption{Exploratory timbre and microtonal examples.}
\label{tab:supp-timbre}
\resizebox{\columnwidth}{!}{
\begin{tabular}{lrr@{\qquad}lrr}
\toprule
Audio & Duration (s) & $D_{\max}$ & Audio & Duration (s) & $D_{\max}$ \\
\midrule
Saxophone & 14.005 & .003134 & Violins & 15.531 & .005633 \\
Guitar & 14.005 & .003170 & Alto & 15.214 & .007333 \\
Piano & 14.099 & .004385 & Trumpet & 14.005 & .011913 \\
Flute & 15.724 & .029903 & 24-TET sequence & 26.005 & .003751 \\
\bottomrule
\end{tabular}}
\end{table}

\begin{figure}[H]
\centering
\includegraphics[width=\columnwidth]{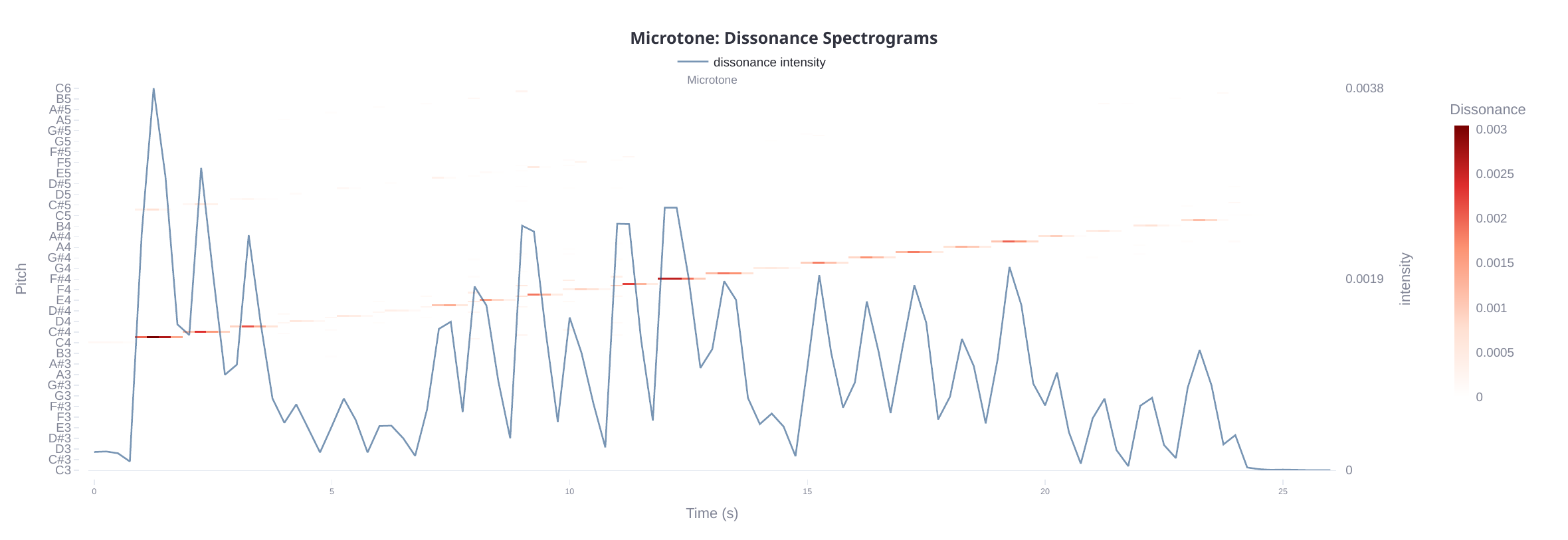}
\caption{Exploratory 24-TET spectra at 50-cent resolution. The timbre values in Table~\ref{tab:supp-timbre} are descriptive and are not assigned a universal ranking.}
\label{fig:supp-timbre-microtone}
\end{figure}

\subsection{Loudness Disentanglement under the Normalized Pipeline}
The primary quantitative test uses seven independently rendered C4--C$\sharp$4 dyads spanning the configured level conditions ($n=7$). Let $L_i=\sum_n|y_i[n]|$ be the absolute-amplitude level proxy used by the validation suite (not a perceptual loudness measure), and let $Y_i$ be the DS peak. For positive $L_i$ and $Y_i$, we fit the robust log--log relation
\begin{align}
\ln Y_i &= a+\beta\ln L_i+\epsilon_i,\\
D_{\mathrm{elastic}} &= \max(0,1-|\beta|),
\end{align}
where $\beta$ is the Theil--Sen slope and a larger $D_{\mathrm{elastic}}$ indicates lower proportional sensitivity. The estimate is $\beta=-.000$ with a Theil--Sen 95\% CI of $[-.014,.000]$, giving $D_{\mathrm{elastic}}=1.000$ after rounding. The DS peak varies by only .94\% relative standard deviation and 2.49\% relative range, defined as $s_Y/\bar Y$ and $(\max_iY_i-\min_iY_i)/\bar Y$, respectively. Thus even the lower confidence bound corresponds to approximately a .014\% DS change per 1\% loudness change. Pearson $r=-.744$ ($p=.0551$) is also reported, but it describes the ordering of small residual deviations rather than their proportional magnitude; a sizeable $|r|$ can therefore coexist with near-zero elasticity. Because the controlled target and reference contexts are normalized, this sanity check supports near gain-invariance of the normalized DS summary over the tested range rather than universal statistical independence from perceptual loudness.

The companion crescendo stimulus visualizes the time-resolved behavior within one file. It is not included in the seven-sample elasticity estimate because adjacent events share a rendering and temporal context. Future tests should expand the number of independent levels and instruments, report LUFS and clipping diagnostics, and repeat rendering under multiple gain and normalization conventions.

\begin{figure}[H]
\centering
\includegraphics[width=.86\columnwidth]{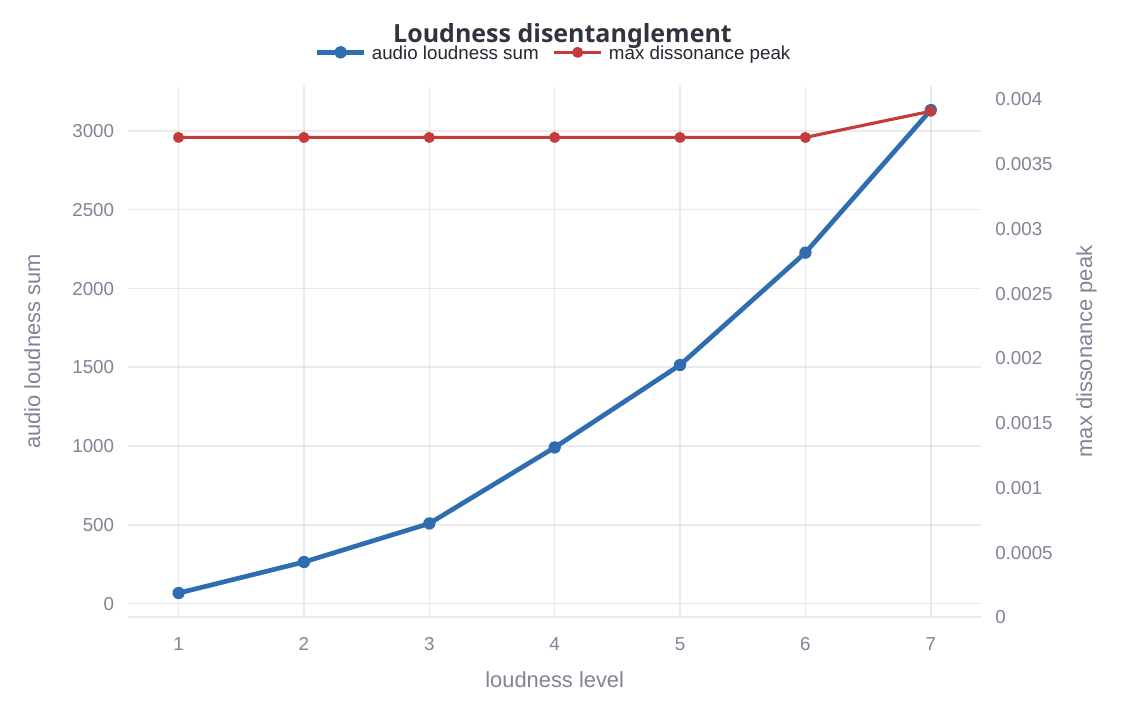}
\caption{Primary seven-level result: the absolute-amplitude level proxy changes substantially while the maximum DS peak remains tightly concentrated.}
\label{fig:supp-loudness}
\end{figure}

\begin{figure}[H]
\centering
\includegraphics[width=.49\columnwidth]{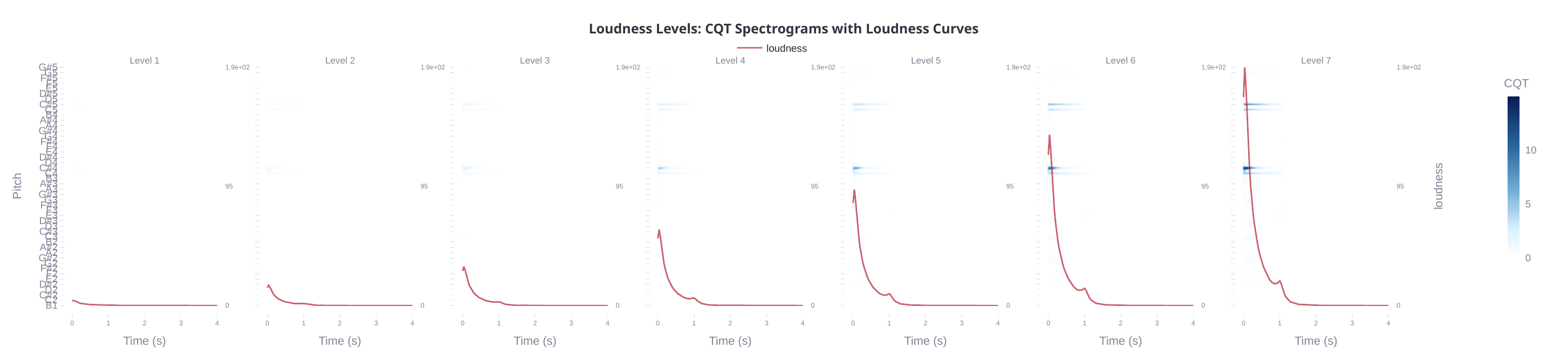}\hfill
\includegraphics[width=.49\columnwidth]{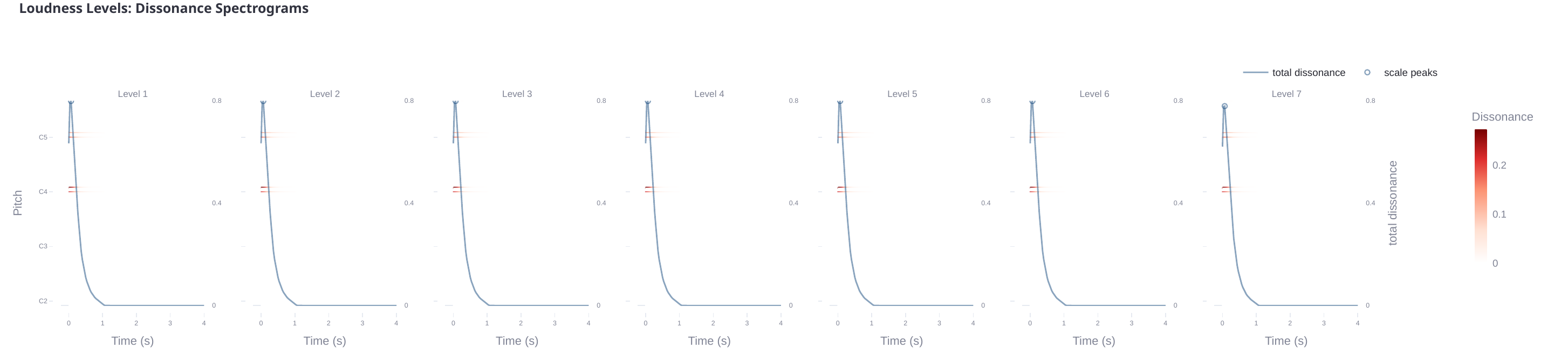}
\caption{Level-controlled stimuli. Left: CQT magnitude. Right: intrinsic DS.}
\label{fig:supp-loudness-level-spectra}
\end{figure}

\begin{figure}[H]
\centering
\includegraphics[width=.49\columnwidth]{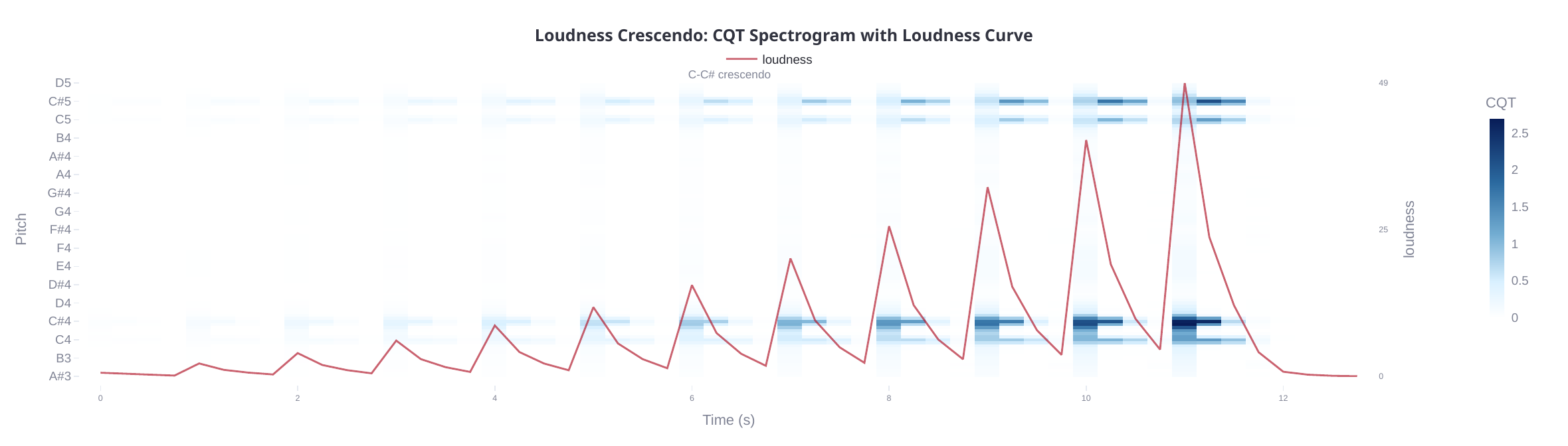}\hfill
\includegraphics[width=.49\columnwidth]{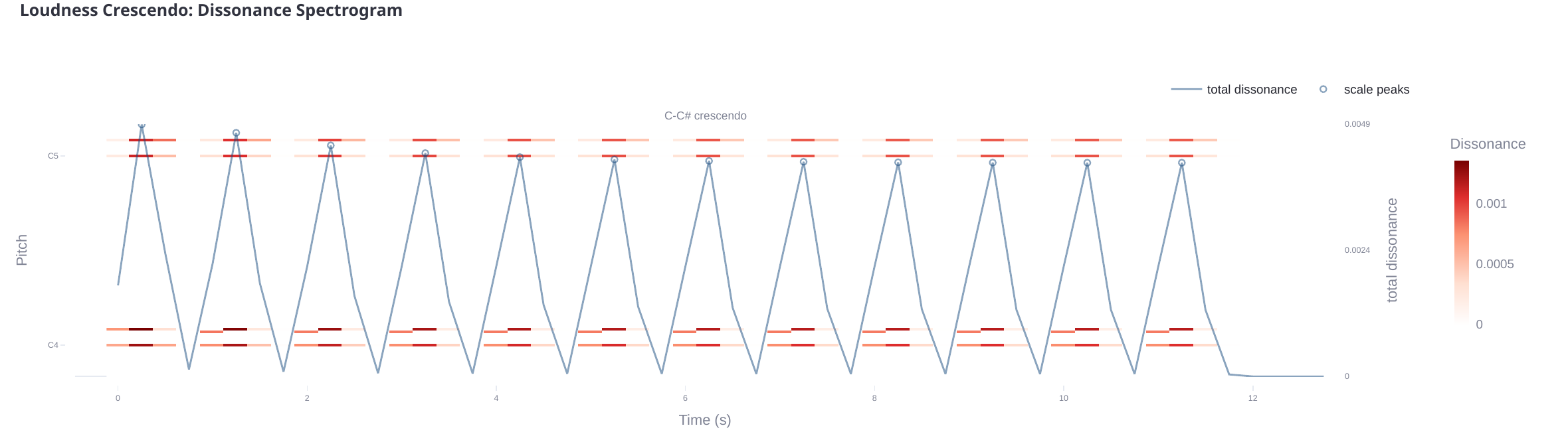}
\caption{Within-file crescendo visualization. Left: CQT magnitude. Right: intrinsic DS.}
\label{fig:supp-loudness-crescendo}
\end{figure}

\FloatBarrier

\section{Detailed Configuration}
\paragraph{Compute and software.}
All downstream runs use one NVIDIA A100 GPU per Slurm job on Linux; no multi-GPU parallelism is used. The implementations are in PyTorch. CQT and DS extraction is cached before training, and the environment lock file in the code archive records exact framework, CUDA, evaluation-package, and pretrained-model revisions.

\paragraph{Shared preprocessing.}
Except for PMEmo's released chorus clips, all branches use 24-kHz, 45-second excerpts. Magnitude CQT uses a 1,024-sample hop, eight octaves, 72 bins per octave, $f_{\min}=\mathrm{C1}$, and excerpt-level global-maximum normalization. DS is nonnegative and uses $\log(1+\mathbf D)$ compression after the relation transform. CQT and DS use the same 576-bin grid and identical 576-to-144 pooling, temporal masks, encoder, fusion location, parameter count, optimizer, schedule, and checkpoint selection. The comparison therefore matches architecture and spectral grid but does not isolate the relation transform from branch-input compression and normalization. Fixed Gaussian inputs are generated once per excerpt and training seed and reused across epochs.

\paragraph{MU-LLaMA parameters.}
The Baseline contains 4,205,568 trainable parameters. The matched parallel branch adds 1,360,193 parameters, yielding 5,565,761 trainable parameters in Gaussian, CQT, and DS. The branch uses three convolutional blocks, a 128-dimensional temporal encoder, a kernel-3 depthwise temporal convolution, single-head temporal attention, and gated residual fusion. The output projection is zero-initialized and the gate bias is $-3$.

\paragraph{Music2Emo parameters.}
The Baseline task network contains 1,071,617 trainable parameters. The parallel encoder and gated cross-attention residual module add 209,345 parameters, yielding 1,280,962 parameters in Gaussian, CQT, and DS. The added branch is 0.22\% of the frozen 95M-parameter MERT encoder and 19.5\% of the trainable task network. Following the Music2Emo 70/15/15 protocol \citep{kang2025unifiedmusicemotionrecognition}, we generate the track-level split once and keep it fixed across all conditions: 1,261/271/270 for DEAM, 495/124/125 for EmoMusic, and 536/116/115 for the 767-track labeled PMEmo subset. These are experiment-specific splits, not official dataset splits. PMEmo contains 794 items in total and uses the released chorus clip.

\section{Evaluation Implementation}
MusicQA uses the archive's explicitly specified evaluator, which differs from the released MU-LLaMA scoring script. NLTK word-punctuation tokenization is applied before sentence BLEU with weights $(.25,.25,.25,.25)$ and method-1 smoothing. METEOR uses exact, stem, and WordNet synonym matching. ROUGE-L uses \texttt{rouge-score} with stemming and reports F-measure. BERTScore uses \texttt{roberta-large} in English without IDF or baseline rescaling and reports recall. Loss covers answer tokens; perplexity is exponentiated per seed before averaging. For MTG-Jamendo, PR-AUC and ROC-AUC are macro-averaged over the 56 tags. The Music2Emo weighted binary cross-entropy uses $w_i=2/(1+p_i)$ for the positive term and $\bar w_i=2p_i/(1+p_i)$ for the negative term, where $p_i$ is the training prevalence of tag $i$. The lock file records exact package and model revisions.

\section{Additional MusicQA Comparisons}
\label{sec:supp-qa-comparisons}
The MusicQA references and model outputs are free-form text. To avoid cherry-picking fluent examples, we summarize additional corpus-level comparisons rather than selecting answers by visual inspection. The questions follow the MU-LLaMA data-generation setting and cover attributes such as mood, instrumentation, tempo, genre, and overall tone \citep{liu2024mullama}. Table~\ref{tab:supp-qa-deltas} reports the difference between the DS condition and each matched control using the six-seed means from the fixed 5,040-pair evaluation set. Higher is better for text-similarity metrics and lower is better for loss and perplexity. Seed-level predictions, questions, and references are included in the separately submitted archive for answer-level inspection.

\begin{table}[H]
\centering
\scriptsize
\caption{Additional MusicQA comparisons. Entries are DS minus the indicated control; negative values are improvements for loss and perplexity.}
\label{tab:supp-qa-deltas}
\setlength{\tabcolsep}{4.0pt}
\begin{tabular}{lrrr}
\toprule
Metric & vs. Baseline & vs. Gaussian & vs. CQT \\
\midrule
BLEU $\uparrow$ & +.0087 & +.0089 & +.0018 \\
METEOR $\uparrow$ & +.0096 & +.0098 & +.0019 \\
ROUGE-L $\uparrow$ & +.0115 & +.0117 & +.0028 \\
BERTScore-R $\uparrow$ & +.0072 & +.0074 & +.0028 \\
Test loss $\downarrow$ & -.025 & -.026 & -.007 \\
Perplexity $\downarrow$ & -.046 & -.048 & -.014 \\
\bottomrule
\end{tabular}
\end{table}

The Gaussian comparison shows that the improvement is not explained by the added parameter budget alone. The smaller but consistently favorable mean difference over the architecture-matched CQT branch indicates that the pitch-resolved input accounts for part, but not all, of the gain. These comparisons remain reference-similarity analyses: automatically generated references can reward paraphrase overlap and do not by themselves establish factual correctness or expert-level harmonic reasoning. We therefore treat the qualitative prediction files as audit material and the paired BERTScore-R endpoint as the prespecified aggregate comparison.

\paragraph{Answer-level audit examples.}
Tables~\ref{tab:supp-qa-index}--\ref{tab:supp-qa-cases-b} reproduce all ten rows in the supplied audit set rather than selecting a fluent subset. Table~\ref{tab:supp-qa-index} maps each case to the audio identifier and MTG-Jamendo track index recorded by the accompanying demo. The \emph{Original Token F1} column is the token-overlap score between the original MU-LLaMA output and the reference answer; it is not a score for the DS-augmented output. Tables~\ref{tab:supp-qa-cases-a} and~\ref{tab:supp-qa-cases-b} provide the corresponding questions, reference answers, and both inference outputs. Relative to the original outputs, the DS-augmented outputs tend to replace generic genre labels, inferred lyric narratives, or vague ``weird'' descriptions with more specific acoustic, stylistic, and affective attributes. These examples are qualitative audit material and do not replace the aggregate paired evaluation.

\begin{table*}[t]
\centering
\footnotesize
\setlength{\tabcolsep}{5pt}
\renewcommand{\arraystretch}{1.08}
\caption{Provenance and score ownership for the MusicQA audit examples. Audio ID and dataset track index reproduce the \texttt{audio} and \texttt{trackId} fields in the accompanying demo. Original Token F1 applies only to the original MU-LLaMA output.}
\label{tab:supp-qa-index}
\begin{tabular}{clrc@{\hspace{1.5em}}clrc}
\toprule
ID & Audio ID & Dataset track index & Original Token F1 & ID & Audio ID & Dataset track index & Original Token F1 \\
\midrule
01 & \texttt{012895.wav} & 661609 & .000 & 06 & \texttt{013055.wav} & 873804 & .071 \\
02 & \texttt{012815.wav} & 477309 & .038 & 07 & \texttt{012553.wav} & 8308   & .111 \\
03 & \texttt{012725.wav} & 267904 & .063 & 08 & \texttt{012986.wav} & 796003 & .073 \\
04 & \texttt{012786.wav} & 387501 & .087 & 09 & \texttt{012972.wav} & 768506 & .094 \\
05 & \texttt{012750.wav} & 345903 & .000 & 10 & \texttt{013063.wav} & 875005 & .103 \\
\bottomrule
\end{tabular}
\end{table*}

\begin{table*}[t]
\centering
\scriptsize
\setlength{\tabcolsep}{2.5pt}
\renewcommand{\arraystretch}{1.08}
\caption{MusicQA answer-level audit examples, cases 01--05. Reference answers and dataset identifiers are reproduced from the accompanying demo.}
\label{tab:supp-qa-cases-a}
\begin{tabular}{p{.025\textwidth}p{.17\textwidth}p{.315\textwidth}p{.18\textwidth}p{.25\textwidth}}
\toprule
ID & Question & Reference answer & Original MU-LLaMA & DS-augmented MU-LLaMA \\
\midrule
01 & What is the term used to describe a style of music that features a strong emphasis on melody and harmony? & Acoustic rock is a term used to describe a style of music that features a strong emphasis on melody and harmony. Acoustic rock typically combines elements of rock and folk music, with acoustic guitar, harmonica, and vocals as the main instruments. & Classical. & Acoustic rock. \\
02 & What type of orchestral music is characterized by complex harmonies and intricate compositions? & Progressive orchestral. Progressive orchestral music incorporates elements of progressive rock, avant-garde, and experimental music, and is characterized by complex harmonies and intricate compositions. & Symphonic music. & Progressive orchestral music. \\
03 & What can be inferred from the audio? & The audio blends folk, punk rock, rock, and ska. It uses acoustic guitars, bass, drums, and occasional brass and strings, with gruff vocals, harmonies, and call-and-response sections. & The song is about a man's love for a woman. & The music blends folk, punk rock, ska, and acoustic instrumentation. \\
04 & Describe the audio in detail. & The audio is a soothing and upbeat tune with a catchy melody and harmonious vocals. Harp, synthesizer, and piano create an elegant, layered, and enjoyable arrangement. & The music is described as slow and weird. & The music is described as soothing, upbeat, and harmonious. \\
05 & What is the genre of music that is characterized by a slow and smooth rhythm, often played on a saxophone? & Bossanova. It originated in Brazil and is known for its relaxed and romantic sound, often featuring a slow and smooth rhythm. & Jazz. & Bossanova. \\
\bottomrule
\end{tabular}
\end{table*}

\begin{table*}[t]
\centering
\scriptsize
\setlength{\tabcolsep}{2.5pt}
\renewcommand{\arraystretch}{1.08}
\caption{MusicQA answer-level audit examples, cases 06--10 (continued).}
\label{tab:supp-qa-cases-b}
\begin{tabular}{p{.025\textwidth}p{.17\textwidth}p{.315\textwidth}p{.18\textwidth}p{.25\textwidth}}
\toprule
ID & Question & Reference answer & Original MU-LLaMA & DS-augmented MU-LLaMA \\
\midrule
06 & What can be inferred from the audio? & The audio appears to be a movie or TV soundtrack with a moody, atmospheric character. Its slow, mellow melody and layered guitars and piano evoke longing, nostalgia, and introspection. & The song is about love and heartbreak. & The music feels moody, mellow, nostalgic, and introspective. \\
07 & Describe the audio in detail. & A relaxing instrumental composition combining electronic, lounge, and oriental music. Its slow, smooth tempo and minimal instrumentation create a calming atmosphere. & The music is described as having a weird and otherworldly quality. & The music is described as relaxing, smooth, and gently oriental. \\
08 & What can be inferred from the audio? & A relaxing and upbeat trip-hop track for chilling out. Synthesizer, strings, bass, drums, Rhodes, and trumpet create a dreamy, smooth, and sophisticated atmosphere. & The song is about love and heartbreak. & The track is relaxing, upbeat, dreamy, and sophisticated. \\
09 & Describe the audio in detail. & A calm orchestral composition. Acoustic bass, cello, piano, violin, bell, and oboe create a warm, elegant, magical, and serene piece for relaxing. & The music is described as having a weird and otherworldly quality. & The music is described as calm, warm, elegant, and serene. \\
10 & Describe the audio in detail. & A fusion of classical and ambient music with flute and synthesizer, a slow and mellow pace, bass and electric guitar foundation, and soulful background vocals. & The music is described as having a weird and scary atmosphere. & The music is described as slow, mellow, soulful, and ambient. \\
\bottomrule
\end{tabular}
\end{table*}

\section{Exploratory Mechanism Analysis}
The mechanism analyses are secondary and are not used to select the reported architecture. MusicQA uses seeds $\{17,42,101\}$; Music2Emo uses all six paired seeds. Global pooling removes temporal localization, temporal shuffling preserves marginal token statistics while destroying order, pre-projection or early fusion changes the insertion point, and the randomized kernel preserves symmetry and its value distribution while permuting frequency correspondence.

\begin{table}[H]
\centering
\scriptsize
\caption{Exploratory MusicQA mechanism analysis.}
\label{tab:supp-qa-mechanism}
\resizebox{\columnwidth}{!}{
\begin{tabular}{lcc}
\toprule
Variant & BERTScore-R $\uparrow$ & $\Delta$ vs. proposed \\
\midrule
Global DS & .8985$\pm$.0007 & $-.0030$ \\
Randomized kernel & .8992$\pm$.0009 & $-.0023$ \\
Temporal DS, shuffled & .8997$\pm$.0007 & $-.0018$ \\
Temporal DS, pre-projection & .9004$\pm$.0006 & $-.0011$ \\
Temporal DS, proposed & \textbf{.9015$\pm$.0013} & -- \\
\bottomrule
\end{tabular}}
\end{table}

\begin{table}[H]
\centering
\scriptsize
\caption{Exploratory Music2Emo mechanism analysis.}
\label{tab:supp-emotion-mechanism}
\resizebox{\columnwidth}{!}{
\begin{tabular}{lc}
\toprule
Variant & Emotion $\rmacro$ $\uparrow$ \\
\midrule
Global DS & .6524$\pm$.0015 \\
Temporal DS, shuffled & .6550$\pm$.0017 \\
Global DS, early concatenation & .6558$\pm$.0016 \\
Temporal DS, proposed & \textbf{.6589$\pm$.0018} \\
\bottomrule
\end{tabular}}
\end{table}

All perturbations reduce the mean relative to the proposed ordered temporal configuration. These results are consistent with a contribution from frequency correspondence and temporal organization, but they do not isolate a single causal mechanism and support no inferential claim.

\section{Reference Comparisons under Modified Protocols}
\begin{table}[H]
\centering
\small
\caption{Contextual MU-LLaMA comparison. The original values use the released scoring script on 4,500 QA pairs from 500 tracks \citep{liu2024mullama}; our values use the explicitly specified evaluator and are six-seed means on the custom 5,040-pair, 560-track set.}
\label{tab:mullama_repro}
\resizebox{\columnwidth}{!}{
\begin{tabular}{lrrrr}
\toprule
Source & BLEU & METEOR & ROUGE-L & BERTScore-R \\
\midrule
Original report (released script) & .3060 & .3850 & .4660 & .9010 \\
Our baseline (specified metrics) & .2987 & .3761 & .4556 & .8952 \\
Absolute difference & .0073 & .0089 & .0104 & .0058 \\
\bottomrule
\end{tabular}}
\end{table}
The rows are not treated as a direct reproduction comparison because both the evaluation set and metric implementation differ.

\begin{table}[H]
\centering
\scriptsize
\caption{Contextual Music2Emo comparison. The original report uses its 30-second segment augmentation and original training protocol \citep{kang2025unifiedmusicemotionrecognition}; our baseline uses the modified fixed-excerpt protocol described here.}
\label{tab:music2emo_repro}
\resizebox{\columnwidth}{!}{
\begin{tabular}{lcccccccc}
\toprule
Source & J-PR & J-ROC & DEAM-V & DEAM-A & Emo-V & Emo-A & PM-V & PM-A \\
\midrule
Original report & .1543 & .7810 & .5184 & .6228 & .6512 & .7616 & .5473 & .7940 \\
Our baseline & .1539 & .7806 & .5169 & .6209 & .6487 & .7598 & .5451 & .7926 \\
Absolute difference & .0004 & .0004 & .0015 & .0019 & .0025 & .0018 & .0022 & .0014 \\
\bottomrule
\end{tabular}}
\end{table}

\section{Seed-Level Primary Endpoints}
\begin{table}[H]
\centering
\scriptsize
\caption{MusicQA BERTScore-R for the six paired seeds.}
\label{tab:qa_seed}
\resizebox{\columnwidth}{!}{
\begin{tabular}{rcccc}
\toprule
Seed & Baseline & Gaussian & CQT & DS \\
\midrule
17 & .8943 & .8942 & .8990 & .9001 \\
42 & .8950 & .8948 & .8996 & .9017 \\
101 & .8954 & .8951 & .8998 & .9027 \\
2025 & .8948 & .8947 & .8989 & .9028 \\
2026 & .8962 & .8960 & .8997 & .9024 \\
3407 & .8955 & .8952 & .9006 & .9047 \\
\midrule
Mean & .8952 & .8950 & .8996 & \textbf{.9024} \\
SD & .0007 & .0006 & .0006 & .0015 \\
\bottomrule
\end{tabular}}
\end{table}

\begin{table}[H]
\centering
\scriptsize
\caption{Music2Emo $\rmacro$ for the six paired seeds.}
\label{tab:emotion_seed}
\resizebox{\columnwidth}{!}{
\begin{tabular}{rcccc}
\toprule
Seed & Baseline & Gaussian & CQT & DS \\
\midrule
17 & .6453 & .6449 & .6519 & .6560 \\
42 & .6472 & .6466 & .6542 & .6587 \\
101 & .6481 & .6476 & .6552 & .6596 \\
2025 & .6464 & .6457 & .6525 & .6580 \\
2026 & .6494 & .6488 & .6557 & .6602 \\
3407 & .6476 & .6468 & .6555 & .6609 \\
\midrule
Mean & .6473 & .6467 & .6542 & \textbf{.6589} \\
SD & .0014 & .0014 & .0016 & .0018 \\
\bottomrule
\end{tabular}}
\end{table}

\section{Paired Comparisons}
\begin{table}[H]
\centering
\scriptsize
\caption{Paired comparisons on the two designated primary endpoints using unrounded seed-level values. $p_{\mathrm H}^{(t)}$ is Holm-adjusted paired-$t$; $p_{\mathrm H}^{(\mathrm{sign})}$ is the Holm-adjusted exact two-sided sign test based only on paired directions. Each task has one three-comparison Holm family.}
\label{tab:paired}
\resizebox{\columnwidth}{!}{
\begin{tabular}{llrrr}
\toprule
Task & Comparison & Mean $\Delta$ & $p_{\mathrm H}^{(t)}$ & $p_{\mathrm H}^{(\mathrm{sign})}$ \\
\midrule
MusicQA & DS vs. Baseline & +.0072 & $<.001$ & .0938 \\
MusicQA & DS vs. Gaussian & +.0074 & $<.001$ & .0938 \\
MusicQA & DS vs. CQT & +.0028 & .0017 & .0938 \\
Music2Emo & DS vs. Baseline & +.0116 & $<.001$ & .0938 \\
Music2Emo & DS vs. Gaussian & +.0122 & $<.001$ & .0938 \\
Music2Emo & DS vs. CQT & +.0047 & $<.001$ & .0938 \\
\bottomrule
\end{tabular}}
\end{table}

\bibliography{references}